\documentclass[11pt,a4paper]{article}
\pdfoutput=1

\usepackage{jheppub}
\usepackage{graphicx}
\usepackage{amsmath}
\usepackage{amssymb}
\usepackage{braket}
\usepackage{hyperref}
\usepackage{comment}
\usepackage{epsfig}
\usepackage{float}
\usepackage{color}
\usepackage{subfigure}
\usepackage[utf8]{inputenc}

\newcommand{\bea}{\begin{eqnarray}}
\newcommand{\eea}{\end{eqnarray}}
\newcommand{\be}{\begin{equation}}
\newcommand{\ee}{\end{equation}}
\newcommand{\ba}{\begin{align}}
\newcommand{\ea}{\end{align}}

\newcommand{\vo}{\mathcal{V}}

\newcommand{\bi}{\begin{itemize}}
\newcommand{\ei}{\end{itemize}}
\newcommand{\ben}{\begin{enumerate}}
\newcommand{\een}{\end{enumerate}}

\newcommand{\F}{\mathcal{F}}

\def\QED{{\scriptscriptstyle \rm QED}}
\def\ULA{{\scriptscriptstyle \rm ULA}}

\def\KK{{\scriptscriptstyle \rm KK}}
\def\W{{\scriptscriptstyle \rm W}}

\def\vo{{{\cal{V}}}}

\newcommand{\mc}{\mathcal}

\def\K3{{\scriptscriptstyle {\rm K3}}}

\def\F4{{\scriptscriptstyle {\rm F}^4}}
\def\cutoff{{\scriptscriptstyle {\rm cutoff}}}

\def\QED{{\scriptscriptstyle {\rm QED}}}
\def\flux{{\scriptscriptstyle {\rm flux}}}
\def\strong{{\scriptscriptstyle {\rm strong}}}

\title{De Sitter vs Quintessence in String Theory}

{\author[1,2,3]{Michele Cicoli,}
\author[4]{Senarath de Alwis,}
\author[5]{Anshuman Maharana,}
\author[3]{Francesco Muia}
\author[3,6]{and Fernando Quevedo}

\affiliation[1]{\small \it Dipartimento di Fisica e Astronomia, Universit\`a di Bologna, via Irnerio 46, 40126 Bologna, Italy}
\affiliation[2]{\small \it INFN, Sezione di Bologna, viale Berti Pichat 6/2, 40127 Bologna, Italy}
\affiliation[3]{\small \it ICTP, Strada Costiera 11, Trieste 34014, Italy}
\affiliation[4]{\small \it Physics Department, University of Colorado, Boulder, CO 80309 USA}
\affiliation[5]{\small \it  Harish Chandra Research Institute, \\
Homi Bhabha National Institute, Jhunsi, Allahabad, Uttar Pardesh, India 211019}
\affiliation[6]{\small \it DAMTP, Centre for Mathematical Sciences, Wilberforce Road, Cambridge, CB3 0WA, UK}

\emailAdd{mcicoli@ictp.it}
\emailAdd{dealwiss@colorado.edu}
\emailAdd{anshumanmaharana@hri.res.in}
\emailAdd{fmuia@ictp.it}
\emailAdd{fq201@damtp.cam.ac.uk}

\abstract{De Sitter solutions have been recently conjectured to be incompatible with quantum gravity. In this paper we critically assess the progress and challenges of different mechanisms to obtain de Sitter vacua in string compactifications and compare them to quintessence models. We argue that, despite recent criticisms, de Sitter models reached a level of concreteness and calculational control which has been improving over time. On the other hand, 
building string models of quintessence appears to be more challenging and requires additional fine-tuning. We discuss the tension between the swampland conjecture and the Higgs potential and find examples which can evade fifth-force bounds even if they seem very hard to realise in string theory. We also comment on the tension with low-redshift data and explore ultra-light axions from string theory as dark energy candidates.}

\begin{document}
\maketitle

\section{Introduction}

Ever since the first superstring revolution, there has been constant progress in the area of string phenomenology~\cite{Ibanez:2012zz, Quevedo:2016tbh}. Based on our current understanding of string theory, the picture of the string landscape with a large number of vacua that can accommodate our universe  (with a positive cosmological constant) has emerged~\cite{Douglas:2006es}. As our understanding of string theory improves and new computational techniques are developed, we should be able to establish the existence of these vacua more and more rigorously and make contact with phenomenology. At the same time, a number of criteria to determine which effective field theories can be consistently embedded into a theory of quantum gravity have been proposed and are called the \textit{swampland conjectures}~\cite{Vafa:2005ui, Ooguri:2006in, Ooguri:2016pdq, Brennan:2017rbf}.  

Effective field theories that can be consistently embedded in string theory are part of the string landscape, as opposed  to  those in the swampland which are not consistent with quantum gravity. The typical example is the swampland conjecture about the boundary of moduli spaces:  any effective field theory is valid only within an $\mc{O}(M_{\rm p})$ field range in field space, since new light states appear in the spectrum of the theory as one moves farther away \cite{Ooguri:2006in}. Recently, a new swampland criterion has been proposed \cite{Obied:2018sgi} which is in contradiction with the picture of a large number of (possibly dS) vacua in the string landscape and inflationary models. The conjecture states that everywhere in field space the full quantum scalar potential $V$ obeys the relation:
\be
M_{\rm p}\,\frac{\left|\nabla V\right|}{V} \gtrsim c \,,
\label{eq:SwamplandConjecture}
\ee
where $c$ is an $\mc{O}(1)$ positive constant. It is important to examine whether such a criterion can be consistent with phenomenology. The criterion (\ref{eq:SwamplandConjecture}) has many strong implications for cosmology \cite{Agrawal:2018own, Andriot:2018wzk, Colgain:2018wgk}. In particular it implies that at present we are necessarily in an epoch of quintessence. The tight bounds on fifth-forces \cite{Adelberger:2003zx} and the time variation of fundamental constants \cite{Martins:2017yxk}, provide strong constraints on the couplings of the quintessence field. Furthermore, in the context of $N=1$ supergravity it seems very hard to be able to decouple a quintessence field from the Standard Model. Finally, depending on the model, naturalness considerations require fine-tuning of the quintessence potential at the functional level\footnote{A similar problem has been discussed in the context of attempts to explain time variation of coupling constants in terms of a time varying field \cite{Banks:2001qc}.}, or at least one additional tuning compared to dS models. This makes explicit constructions of quintessence models from string compactifications very challenging.

This conjecture is the most recent of a series of articles claiming potential problems with the standard approach to obtain a landscape of metastable dS string vacua as initiated by the KKLT seminal paper \cite{Kachru:2003aw} and followed-up by many other developments that have improved the robustness of the original and other related scenarios. The challenges vary from points of principle (e.g. how to properly define an S-matrix and a quantum theory in general in dS space \cite{Witten:2001kn, Banks:2012hx, Maltz:2016iaw}) to details about each of the different steps of the KKLT scenario \cite{Sethi:2017phn, Bena:2009xk, Moritz:2017xto} which seem to make it natural to explore alternatives to dS. The main purpose of the first part of this article is to assess the pros and cons of the different approaches to dS compactifications. This is important in order to have a clear idea of the assumptions used and the continuous progress but also the open challenges. We will argue that dS models reached a good level of concreteness and calculational control which has been improving over time and provide interesting phenomenological applications to cosmology and particle physics. Moreover we shall stress that some of the computational challenges apply also to 4D $N=1$ supersymmetric vacua which, above all, do not seem to be promising starting points for phenomenology. We will also point out that, even if dS string models are not characterised by expansion parameters which can be made parametrically small, these parameters can still be small enough to trust the phenomenological implications of these constructions.

In the second part of the paper we first discuss the theoretical consistency of quintessence models pointing out that in general, in the absence of a symmetry principle, their construction is more challenging that dS models since one needs to perform two fine-tunings to get the correct energy scale and mass of the quintessence field. We then use a more phenomenological approach to assess to which extent quintessence is a viable alternative to dS from observations. In particular, we found (as recently shown also in \cite{Denef:2018etk}), that if the quintessence picture is valid, and there is no other scalar field around other than the Higgs, in order to satisfy the swampland conjecture \eqref{eq:SwamplandConjecture}, the Higgs field has to couple directly to the quintessence field. This would be particularly challenging in any string theory/supergravity scenario, as a quintessence field that couples directly to the Higgs would also couple to the SM fermions, violating fifth-force constraints. We explore these issues, providing examples that avoid the direct coupling of the Higgs to quintessence even if they seem very hard to realise in a supergravity setup.

The paper is organised as follows. Sec. \ref{sec:dSstrings} is devoted to the discussion of dS models from string theory. After briefly recalling the need for dS in Sec. \ref{WhydS}, we provide a review of pros and cons of type IIB string models to achieve dS vacua in Sec. \ref{sec:ProsCons}. We then turn to a more detailed analysis of various advantages and criticisms of dS vacua from anti-branes in Sec. \ref{Dbar} and from T-branes in Sec. \ref{T}, while Sec. \ref{OtherdS} contains short comments on other existing string mechanisms to achieve dS vacua. After that, we turn to quintessence in Sec. \ref{sec:QuintessenceStrings}. In particular we discuss various general challenges for quintessence model building in Sec. \ref{sec:QuintessenceChallenges} and the constraints on the coupling between the Higgs and the quintessence field due to the swampland conjecture in Sec. \ref{sec:SwampandHiggs}. We then 
 review the quintessence models already present in the literature in Sec. \ref{sec:StringQuintessenceModels} and finally we study the r\^ole that ultra-light axions can play to explain dark energy data in Sec. \ref{sec:LowRedshift}.

\section{De Sitter in string theory}
\label{sec:dSstrings}

\subsection{Why dS?}
\label{WhydS}

Present observations suggest that the current energy density of dark energy is $\rho_0 \sim 10^{-120}\,M_{\rm p}^4$ and that it is consistent with a positive cosmological constant. A concrete way to quantify this is through the equation of state parameter $w=p/\rho$ which naturally hints towards an asymptotic dS vacuum or something very close to it as recently reported by the Planck collaboration \cite{Aghanim:2018eyx}:
\be
w_0=-1.028\pm 0.032\,.
\ee
If dark energy is described by the vacuum energy, we need a scalar potential $V$ whose size today is of order $\langle V\rangle_0 \sim \rho_0\sim \Lambda^4 \sim ({\rm meV})^4$. Other possibilities involve quintessence models where the vacuum energy might be exactly zero (for example due to some yet to be found symmetry arguments) or negative, while at present the quintessence field is slow-rolling at positive energies. However recent low-redshift data show some tension with $\Lambda$CDM and seem to disfavour quintessence models (for a recent discussion and references see \cite{Wang:2018fng, Capozziello:2018jya, Dutta:2018vmq}).

In a Wilsonian approach, the value of $\langle V\rangle_0$ is the result of integrating out all modes from the UV down to the cosmological constant scale $\Lambda\sim {\rm meV}$. In string models, the 4D Wilsonian effective action is evaluated by integrating out all stringy and Kaluza-Klein modes down to the compactification scale $M_\KK \gg \Lambda$. Hence the vacuum energy computed in a 4D string compactification model $\langle V\rangle$ differs from $\langle V\rangle_0$ since it does not take into account corrections coming from integrating out light degrees of freedom associated to any energy scale $M$ between $M_\KK$ and $\Lambda$. Hence we have $\langle V\rangle_0 = \langle V\rangle + \mc{O}(M^4)$. It is reasonable to expect $M\sim {\rm TeV}$ even if larger energy scales could also be present. Depending on the sign of these corrections, $\langle V\rangle$ can in principle be both positive and negative. It is therefore important to be able to obtain 4D string vacua with a positive vacuum energy which can be tuned to cancel potentially negative low-energy $\mc{O}(M^4)$ corrections and give $\langle V\rangle_0 \sim \Lambda^4$. This is usually guaranteed by the flux landscape where one can choose background fluxes so that $\langle V\rangle$ cancels off any low energy correction, and then focuses on a small perturbation of the flux superpotential which gives $\langle V\rangle_0 \sim \Lambda^4$.

\subsection{Type IIB models: pros and cons}
\label{sec:ProsCons}

Moduli stabilisation is better understood in the context of type II models, and so we will focus only on type II dS constructions (see however \cite{Cicoli:2013rwa} for dS vacua in heterotic models\footnote{See also
\cite{Parameswaran:2010ec} for dS saddle points in heterotic
orbifolds.} and \cite{Acharya:2008zi} for dS solutions in M-theory models on $G_2$ manifolds). Type IIA models allow one to fix all the moduli at tree-level thanks to background fluxes. However so far no stable dS solution has been found \cite{Caviezel:2008tf, Flauger:2008ad, Danielsson:2009ff, Caviezel:2009tu, Danielsson:2010bc, Danielsson:2011au, Andriot:2018ept}. These constructions have the advantage of stabilising the moduli in 10D and at the classical level. However the 10D equations can be solved exactly only under the approximation of smeared sources which would lead to a Calabi-Yau internal manifold. However, in the localised case, the 4D effective field theory picture is not under control since the backreaction of the fluxes on the internal geometry cannot be neglected and leads to a half-flat non-Calabi-Yau metric \cite{Acharya:2006ne, McOrist:2012yc}. This is a serious issue for the trustability of these solutions.

Non-geometric constructions seem to yield dS vacua without tachyons \cite{deCarlos:2009fq, Danielsson:2012by, Blaback:2013ht, Damian:2013dq, Damian:2013dwa}. However also in this case the form of the effective action is not fully under control since the exact form of the moduli space is unknown. Moreover a tree-level stabilisation procedure naturally gives rise to a 4D potential of order the string scale with $\mc{O}(1)$ values of the internal volume, and so it is not clear if $\alpha'$ effects can consistently be neglected.  

Type IIB models are instead characterised by the no-scale structure which makes the K\"ahler moduli massless at tree-level. These directions are then stabilised by the inclusion of perturbative (in both $\alpha'$ and $g_s$) and non-perturbative corrections to the effective action which allow to find stable vacua in the regime of large volume and weak coupling. In this way one can avoid the main part of the Dine-Seiberg problem \cite{Dine:1985he}. For this reason several dS mechanisms have been proposed within the type IIB framework. In what follows we shall first briefly review the main features of type IIB flux compactifications, and then discuss the advantages and challenges of these constructions.

\subsubsection{Overview of IIB flux compactifications}

Type IIB compactifications on orientifolds of a Calabi-Yau (CY) threefold $X$ have several special features that make them promising frameworks to address moduli stabilisation. Let us briefly review the structure of their effective field theory (EFT). The relevant fields are the axio-dilaton $S$, the complex structure moduli $U_a$, $a=1,\cdots, h^{1,2}$ and the K\"ahler moduli $T_i$, $i=1,\cdots, h^{1,1}$ where $h^{1,2}$ and $h^{1,1}$ are the Hodge numbers of the compact CY space. The tree-level K\"ahler potential is:
\be
K=-2\ln\vo-\ln\left(S+\bar{S}\right)-\ln \left(-{\rm i}\, \int_X \Omega\wedge \bar{\Omega}\right),
\label{Ktree}
\ee
where $\vo = \ell_s^{-6} \int_X \sqrt{g_{(6)}}\, d^6 y$ is the CY volume in units of the string length $\ell_s$. The internal volume $\vo$ is a homogeneous function of degree $3/2$ of the real parts of the K\"ahler moduli $\tau_i$ that determine the sizes of internal 4-cycles. $\Omega$ is instead the holomorphic $(3,0)$-form of the CY manifold. In the presence of fluxes, the superpotential takes the form \cite{Gukov:1999ya}:
\be
W_\flux=\int_X G_3\wedge \Omega\,,
\label{Wflux}
\ee
where the 3-form flux $G_3=F_3-{\rm i}SH_3$ contains the NS-NS flux $H_3$ and the RR flux $F_3$. These 3-form fluxes are quantised since their integrals over 
the many 3-cycles of the CY space give rise to flux integers which generate a potential for the $S$ and $U$-moduli. Hence the superpotential (\ref{Wflux}) naturally fixes the dilaton and all complex structure moduli and reduces the number of vacua from a continuum to a discrete but large set of points determined by quantised 3-form fluxes \cite{Dasgupta:1999ss, Giddings:2001yu}. The minimisation conditions require $G_3$ to be imaginary self-dual, i.e. $*_6 G_3 ={\rm i} G_3$, which is compatible with the Hodge decomposition $G_3\in(2,1)\oplus(0,3)$. Notice that in general this solution breaks supersymmetry since supersymmetry is preserved only if the $(0,3)$ component is turned off, as considered in \cite{Dasgupta:1999ss}.  

The K\"ahler moduli $T_i$ are not stabilised by 3-form fluxes. The reason behind this is the fact that there exists a Peccei-Quinn symmetry $T_i\rightarrow T_i +{\rm i} c_i$ with constant $c_i$'s that, together with the holomorphicity of the superpotential, forbids any $T_i$ dependence of $W$ to all orders in perturbation theory. However these moduli are the gauge couplings for matter fields localised on D7-branes, and so effects like gaugino condensation on D7-branes or Euclidean D3-instantons \cite{Blumenhagen:2009qh} generate a non-perturbative superpotential for these fields. The total superpotential for closed string moduli is:
\be
W=W_{\rm flux}(S,U)+W_{\rm np}(S,U,T)\,.
\label{Wtotal}
\ee
The starting point  of the 4D EFT is the F-term supergravity scalar potential for arbitrary superpotential $W(\Phi_M)$ and K\"ahler potential $K(\Phi_M, \bar{\Phi}_{\bar{M}})$ in units of $M_{\rm p}$:
\be
V_F= e^K\left(K^{M\overline{N}}\, D_M W \overline{D}_{\overline{M}} \overline{W} - 3|W|^2\right) \,,
\label{VF}
\ee
where $D_M W = \partial_M W + \left(\partial_M K\right) W$. The tree-level K\"ahler potential for the K\"ahler moduli satisfies the celebrated no-scale property $K^{T_i\bar{T}_{\bar{j}}} K_{T_i} K_{\bar{T}_{\bar{j}}}=3$ which is just a consequence of the homogeneity of $\vo$. Using this and the fact that the flux superpotential does not depend on the $T$-fields, the scalar potential can be easily shown to be positive definite for the $S$ and $U$-moduli which are stabilised supersymmetrically by solving $D_{U_a} W=D_SW=0$. As long as these equations have solutions for different values of the quantised fluxes, they generate the huge number of solutions that define the string landscape. However at this stage all these are Minkowski vacua where the K\"ahler moduli are still exact flat directions. For a generic non-vanishing vacuum expectation value of the tree-level flux superpotential $W_0\equiv W_{\rm flux}(\langle S\rangle,\langle U\rangle)$, corresponding to a non-zero $(0,3)$ component of $G_3$, the $T$-moduli break supersymmetry since $D_{T_i} W = K_{T_i} W_0\neq 0$ \cite{Giddings:2001yu}.

Two main scenarios have emerged to fix the K\"ahler moduli: the original KKLT proposal \cite{Kachru:2003aw} and the Large Volume Scenario (LVS) \cite{Balasubramanian:2005zx, Conlon:2005ki, Cicoli:2008va}. Both focus on the $W_0\neq 0$ case. KKLT uses the fact that $W_0$ can be tuned to small values in order to compete with the small non-perturbative effects in $W_{\rm np}$ to produce an AdS minimum for the $T$-fields. In this case the minimum is at $D_{T_i} W=0$ and supersymmetry is restored. On the other hand, in LVS, instead of tuning $W_0$, the leading order no-scale breaking effect, which is a $\vo$-dependent $\alpha'$ correction, competes with non-perturbative corrections that depend on a blow-up mode wrapped by an ED3-instanton or a D7-stack supporting gaugino condensation. At the resulting AdS minimum, the volume $\vo\sim e^{1/g_s}\gg 1$ is exponentially large in string units and supersymmetry is broken by the F-terms of the K\"ahler moduli.

Since perturbative and non-perturbative effects play an important r\^ole to fix the K\"ahler moduli, let us sketch the general structure of these corrections to $K$ and $W$. First of all, it is crucial to observe that string theory has no free parameter since each coupling corresponds to the value of a different modulus: the string coupling $g_s = 1/{\rm Re}(S)$ is determined by the dilaton which sets also the coupling of gauge theories living on D3-branes at singularities, while the K\"ahler moduli control $\alpha'$ effects which come in an expansion in inverse powers of $\vo$ and the coupling of gauge theories living on D7-branes wrapping internal 4-cycles. Hence stabilising the moduli corresponds to fixing the value of the expansion parameters. Contrary to standard field theories, string compactifications therefore feature many expansion parameters. This makes difficult to extract exact results but also provides much flexibility regarding weak coupling expansions. For weak coupling, the leading order correction to the tree-level K\"ahler potential for the $T$-moduli in (\ref{Ktree}) comes from perturbative effects (either in $\alpha'$ or $g_s$) and we generically denote it as:
\be
K=-2\ln\vo+ K_{\rm p}\,.
\label{Kp}
\ee
The total superpotential (\ref{Wtotal}) takes instead the schematic form:
\be
W=W_0+ W_{\rm np}\,.
\label{Wnp}
\ee
Thus the F-term scalar potential can be expanded as: 
\be
V=V_0 +\delta V\,,
\ee
where the tree-level potential $V_0$ is positive definite due to the no-scale structure and vanishes at the minimum after the $S$ and $U$-moduli are stabilised. In the space of solutions for which ${\rm Re}(S)\gg 1$, the string loop expansion is under control. Since $V_0=0$, the minimum of the potential in the K\"ahler moduli space is determined by the quantum corrections $\delta V$. Determining the leading contributions to $\delta V$ is therefore crucial to properly stabilise the moduli. From the expansions in (\ref{Kp}) and (\ref{Wnp}), the structure of $\delta V$ takes schematically the form \cite{Conlon:2005ki}:
\be
\delta V\propto e^K\left( W_0^2 \,K_{\rm p} + W_0\, W_{\rm np}\right)\,.
\label{eq:deltaV}
\ee 
If there were only one single expansion parameter and if $W_0\gg W_{\rm np}$ and $K_{\rm p}\gg W_{\rm np}$ (since at weak coupling perturbative physics dominates over non-perturbative terms), the first term would be the leading order term. It would lift the potential but would give rise to a runaway behaviour, unless terms of different order in the perturbative expansion compete to give a minimum which would however arise only in a regime where the perturbative expansion breaks down since the corresponding expansion parameter would not be small. This is the Dine-Seiberg problem \cite{Dine:1985he}. 

Type IIB flux compactifications provide two ways to overcome this problem. First, in the KKLT scenario the big discrete degeneracy of flux vacua is used to tune $W_0$ to an exponentially small value so that $W_0\sim W_{\rm np}$. This then requires $W_{\rm np}^2$ terms to be also included in (\ref{eq:deltaV}) stabilising the $T$-fields when they compete with $W_0\, W_{\rm np}$ terms \cite{Kachru:2003aw}. Notice that in this limit quantum corrections to the K\"ahler potential can be consistently neglected since the first term in (\ref{eq:deltaV}) is subdominant given that $W_0^2 \,K_{\rm p} \ll W_0\, W_{\rm np} \sim W_0^2$ for $K_{\rm p}\ll 1$ (this is always the case at large volume since the perturbative effects $K_{\rm p}$ are suppressed by inverse powers of $\vo$).

The second case is LVS models where the fact that there is more than one expansion parameter plays the key r\^ole. In this case the two terms in (\ref{eq:deltaV}) can compete with each other to provide a minimum as long as each comes from a different expansion. Hence at the minimum one has $W_0^2\,K_{\rm p}\sim W_0 \,W_{\rm np}$ which, for $K_{\rm p}\sim 1/\vo$ and $W_{\rm np}\sim e^{-{\tau_s}}$, yields an overall volume of order $\vo\sim W_0\, e^{\tau_s}$. Here $\tau_s$ is a blow-up mode that gets stabilised to values of order $1/g_s$. It is therefore large for weak string coupling, implying that the CY volume is exponentially large \cite{Balasubramanian:2005zx, Conlon:2005ki, Cicoli:2008va}. 

In summary, KKLT requires a major tuning of the fluxes to obtain $W_0\sim W_{\rm np}\ll 1$, whereas LVS works for natural values of the flux superpotential of order $W_0\sim \mc{O}(1-100)$ (as found in concrete examples \cite{Louis:2012nb, Cicoli:2013cha}) but depends more on perturbative corrections to $K$. Notice that, from the $e^K$ factor in the general expression (\ref{VF}), the order of $V_0$ is $V_0\sim M_{\rm p}^4/\vo^2 \sim M_s^4$, whereas in LVS the order of $\delta V$ is $\delta V\sim W_0^2 M_{\rm p}^4 /\vo^3 \sim M_s^2 m_{3/2}^2\ll M_s^4$. Having $V_0$ vanishing at the minimum and $\delta V\ll M_s^4$ supports the  validity of the EFT at scales below $M_s$.

\subsubsection{Advantages}

We would like here to emphasise several advantages of type IIB constructions:
\ben
\item \emph{Controlled flux backreaction}: Background fluxes can be turned on to generate a potential for the moduli in a controlled way since their backreaction on the internal geometry just renders the compactification manifold conformally Calabi-Yau. Therefore the understanding of the underlying moduli space is better than in other string theories. Some progress has been  made  recently in computing the  form of the K\"ahler potential including the effect of warping \cite{deAlwis:2003sn,Giddings:2005ff,Douglas:2007tu,Shiu:2008ry,Douglas:2008jx,Frey:2008xw,Martucci:2009sf,Martucci:2014ska}. Notice that the warping induces corrections to the definition of the correct moduli coordinates which are however negligible at large volume. 

\item \emph{Suppressed scalar potential scale}: The starting point of dS models is the classical low-energy limit of type IIB string theory compactified on a CY orientifold. This is a controlled procedure if the compactification volume is large so that the following hierarchy of scales is valid:
\be
E\ll M_\KK=\frac{M_s}{\vo^{1/6}}\ll M_s \equiv\frac{1}{\ell_s} \equiv \frac{1}{2\pi\sqrt{\alpha'}} =g_s^{1/4}\frac{M_{\rm p}}{\sqrt{4\pi\vo}}\,.
\ee
As mentioned above, at tree-level the dilaton and the complex structure moduli are fixed supersymmetrically at $D_S W=D_U W=0$ via non-zero quantised $G_3$ fluxes \cite{Dasgupta:1999ss, Giddings:2001yu}. On the other hand, the K\"ahler moduli remain flat directions due to the no-scale structure. The scale of the potential at tree-level is of order $V_0 \sim M_s^4$ but its vacuum energy is vanishing due to the no-scale cancellation. This cancellation allows one to keep the value of the scalar potential around this minimum below the string and the Kaluza-Klein scale, and so guarantees that the effective field theory approach is under control.

\item \emph{Suppressed SUSY breaking scale}: As explained above, the
minimisation conditions $D_S W=D_U W=0$ imply that $G_3$ can only have
$(2,1)$ and $(0,3)$ components. Hence in general supersymmetry is broken
at tree-level by the F-terms of the K\"ahler moduli which are proportional
to the $(0,3)$ component of $G_3$ and scale as $F^T =
e^{K/2}\,K^{T\bar{T}} K_{\bar{T}} W_0 \sim \frac{W_0}{\vo^{1/3}}$.
Therefore the scale of supersymmetry breaking is very low since the
gravitino mass $m_{3/2}= e^{K/2} W_0 \sim \frac{W_0}{\vo}$ is
hierarchically smaller than the Kaluza-Klein scale $M_{\rm KK}\sim
M_s/\vo^{1/6} \sim 1/\vo^{2/3}$ for either $W_0\ll 1$ (as in KKLT
constructions) or $\vo \gg 1$ (as in LVS models)\footnote{Notice that in
F-theory models where the string coupling can be arbitrarily large, this
tree-level analysis clearly cannot be trusted.}. Thanks to this
suppression of the supersymmetry breaking scale, it is thus sensible to
compute non-perturbative corrections to the superpotential in a
supersymmetric setup even if ref. \cite{Sethi:2017phn} pointed out that
this can be rigorously done only in the specific case where only $(2,1)$
background fluxes are turned on as considered in \cite{Dasgupta:1999ss}.
In fact, in this case $W_0=0$ which implies $F^T=0$. However this case
necessitates a purely non-perturbative stabilisation of the $T$-moduli
which requires a racetrack-type superpotential whose microscopic origin is
only poorly understood.

\item \emph{Absence of quantum instabilities}: The inclusion of $\alpha'$ corrections to $K$ leads to a runaway instability for the volume mode in the limit where the string coupling is set to zero. This has been claimed to be a potential problem in \cite{Sethi:2017phn}. However systems which are classically unstable do not need to be necessarily unstable also at the quantum level. In fact, when $g_s$ effects are turned on, non-perturbative corrections to the superpotential can dynamically turn out to be of the same order as $\alpha'$ effects in the regime of exponentially large volume where the scale of supersymmetry breaking is very small compared to the string scale. This is the case of LVS models where an analysis which includes only $\alpha'$ corrections but not instanton effects would be inconsistent since the stabilisation procedure shows that these two effects are of the same order of magnitude \cite{Balasubramanian:2005zx, Conlon:2005ki, Cicoli:2008va}. Notice that this scenario works for the generic case when there is more than one K\"ahler modulus. It is precisely this feature that makes the scenario work since a minimum (which in the simplest case without additional sectors responsible to achieve dS is non-supersymmetric AdS) is dynamically generated by the competition of two \textit{different} expansions: the perturbative $\alpha'$ expansion in powers of $1/\vo$, and the non-perturbative expansion for the small modulus in $e^{-\tau_s}$. This explains the exponentially large volume $\vo\sim e^{1/g_s}$ since $\tau_s\sim 1/g_s$ where $g_s$ is taken to be in the weak coupling regime after dilaton stabilisation by suitable 3-form fluxes. On the other hand, in KKLT models, the tuning of the flux superpotential $W_0$ to small values (assuming it can be done for which a large number of complex structure moduli is usually needed) renders non-perturbative corrections to the 4D scalar potential even dominant with respect to $\alpha'$ contributions which can therefore be safely neglected. In this case a supersymmetric AdS vacuum is obtained by balancing $W_0$ against non-perturbative effects \cite{Kachru:2003aw}.

\item \emph{Progress in computing quantum effects}: A lot of progress has been made during the last years to compute non-perturbative contributions to the superpotential (Euclidean D3-brane instantons in particular \cite{Blumenhagen:2009qh}) and perturbative (both in $\alpha'$ and $g_s$) corrections to the K\"ahler potential. After the original computation of $N=2$ $\mc{O}(\alpha'^3)$ corrections to the K\"ahler potential $K$ \cite{Becker:2002nn}, additional $N=2$ $\mc{O}(g_s^2 \alpha'^2)$ and $\mc{O}(g_s^2 \alpha'^4)$ contribution to $K$ have been derived in \cite{Berg:2005ja} and generalised in \cite{Berg:2007wt}. Ref. \cite{Cicoli:2007xp} showed the existence of an \textit{extended no-scale structure} since $\mc{O}(g_s^2 \alpha'^2)$ contributions to the scalar potential cancel off. This result is crucial for the stability of LVS models. Relatively recently there has been substantial progress in understanding also $N=1$ perturbative effects. Ref. \cite{Grimm:2013gma} showed that $N=1$ $\mc{O}(\alpha'^2)$ corrections to the effective action give rise to moduli redefinitions, while ref. \cite{Minasian:2015bxa} found that $N=1$ $\mc{O}(\alpha'^3)$ effects produce a shift of the CY Euler number term\footnote{See also \cite{Anguelova:2010ed} for $N=1$ $\mc{O}(\alpha'^2)$ corrections to $K$ in heterotic constructions which should get mapped to type IIB $\mc{O}(g_s^2\alpha'^2)$ effects that enjoy the extended no-scale cancellation.}. Moreover, ref. \cite{Bonetti:2016dqh} reconsidered $N=2$ $\mc{O}(\alpha'^3)$ contributions to $K$ including the backreaction of these corrections on the internal geometry and found that they induce moduli redefinitions. Interesting progress has also been made in the computation of higher derivative $N=2$ $\mc{O}(\alpha'^3)$ terms \cite{Ciupke:2015msa, Grimm:2017okk} which can have promising implications for moduli stabilisation and cosmology \cite{Broy:2015zba, Cicoli:2016chb}. Finally ref. \cite{Berg:2014ama, Haack:2015pbv,Haack:2018ufg} have recently derived $N=1$ string loop corrections to the Einstein-Hilbert term showing that they generate $g_s^2$ corrections to a term involving the CY Euler number\footnote{See also \cite{Antoniadis:2018hqy} for additional $g_s$ corrections to $K$.}. We list the important corrections that still remain to be computed in the next section focused on challenges.

It is worth stressing that none of the perturbative $\alpha'$ and $g_s$ corrections listed above create instabilities for LVS models. On the other hand, corrections which are subleading in an inverse $\vo$-expansion turn out to be very useful to lift leading order flat directions with interesting implications for cosmology and particle phenomenology. Notice also that sometimes one does not need to derive the full functional dependence of these corrections on all moduli, but it is sufficient to determine their dependence on the K\"ahler moduli which have still to be stabilised. Moreover, the functional dependence of string loop corrections to $K$ on the K\"ahler moduli is the easiest to determine (together with the dilaton dependence) from both generalisations of toroidal computations and low-energy arguments \cite{Cicoli:2007xp, Burgess:2010sy}. Another powerful tool is the requirement of the positivity and convergence of the K\"ahler metric (see section 5.2 of \cite{Conlon:2006gv}). On the other hand, the dependence on the $U$-moduli is the hardest to determine but, given that the complex structure moduli have already been fixed at tree-level in terms of background fluxes, these can be safely considered just to give rise to tunable $\mc{O}(1)$ coefficients. 

Furthermore, it has been established in \cite{Burgess:2005jx} that even though the flux superpotential $W_0$ depends explicitly on the dilaton which is directly related to the string coupling, the superpotential is still not renormalised at any order in perturbation theory. This is non-trivial since the standard arguments for the non-renormalisability of $W$ relied on the fact that $W$ did not depend on the string coupling \cite{Witten:1985bz, Burgess:1985zz, Dine:1986vd}.
 
\item \emph{Controlled higher derivative corrections}: As shown in \cite{Cicoli:2013swa}, the superspace derivative expansion is under control if $W_0\ll \vo^{1/3}$ which corresponds to requiring a gravitino mass which is hierarchically smaller than the Kaluza-Klein scale. This can be guaranteed by either tuning $W_0\ll 1$ as in KKLT models or by $\vo \gg 1$ as in LVS constructions.

\item \emph{Hierarchies for phenomenology}: Type IIB models where non-perturbative effects play a crucial r\^ole for moduli stabilisation are particularly promising for phenomenological applications. In fact, they can generate hierarchies exploiting the exponential suppression typical of non-perturbative effects. This allows one to obtain energy scales like the inflationary scale, the gravitino mass, the soft terms or the moduli masses which are much smaller than the string scale. Without using quantum effects, it is at the moment unknown how to obtain nice phenomenological implications of string vacua.

\item \emph{dS mechanisms}: Several mechanisms have been proposed to obtain dS vacua in type IIB models. In this paper we will avoid the use of the terminology \emph{uplift} since it conveys the wrong idea that moduli stabilisation proceeds in two steps, obtaining first an AdS vacuum which is subsequently uplifted to dS by adding by hand a new ingredient in the compactification. The mechanisms proposed in the literature proceed instead in just a single step where a dS vacuum is achieved by the interplay of several contributions to the 4D scalar potential. Some of the most popular dS mechanisms are: ($i$) anti-branes \cite{Kachru:2003aw}, ($ii$) T-branes \cite{Cicoli:2015ylx}, ($iii$) $\alpha'$ effects \cite{Westphal:2006tn}, ($iv$) non-perturbative effects at singularities \cite{Cicoli:2012fh}, ($v$) non-zero $S$ and $U$ F-terms \cite{Gallego:2017dvd}. 

\item \emph{Explicit global models}: A fully working 4D string model, should not lead just to a dS vacuum but it should also include SM-like chiral matter, an inflationary sector and a concrete embedding in globally consistent Calabi-Yau compactifications with an explicit choice of the orientifold involution, the brane setup, background and gauge fluxes. A lot of progress in this direction has been made recently within the type IIB framework \cite{Cicoli:2011qg, Cicoli:2012vw, Cicoli:2013mpa, Cicoli:2013cha, Cicoli:2016xae, Cicoli:2017shd, Cicoli:2017axo}.

\item \emph{Freedom to tune the vacuum energy}: As explained above, obtaining a dS vacuum is not sufficient to match observational data since one should also have enough tuning freedom to reproduce $\langle V\rangle_0 = \Lambda^4$. In type IIB scenarios, contrary to type IIA or non-geometric constructions, this is guaranteed by the fact that the number of flux quanta (from RR and NSNS 3-form fluxes) is twice as large as the number of moduli fixed at tree-level. Notice also that the main phenomenological features of a given model are almost insensitive to this tuning of the cosmological constant. 

\item \emph{Sources of open string moduli fixing}: Most of the dS constructions available in the literature just focus on the stabilisation of the closed string moduli but a full working model should include the stabilisation of the open string moduli as well. For a concrete global model with full closed and open string moduli stabilisation see \cite{Cicoli:2017axo}. In type IIB constructions with D3 and D7 branes most of the open string moduli get fixed by background fluxes. These can be seen as supersymmetry breaking soft term contributions to the scalar potential of D7 deformation moduli and open strings at the intersection between different stacks of D7-branes \cite{Camara:2004jj}. D3 open string modes are instead flat directions at tree-level but they can be stabilised by non-zero soft term masses which can arise either from $\alpha'$ corrections to the matter K\"ahler metric and non-zero F-terms of the $T$-moduli induced by 3-form fluxes, or from non-zero F-terms of the $S$ and $U$-moduli corresponding to IASD background fluxes which are dynamically induced by quantum corrections to the GKP solution \cite{Aparicio:2014wxa}. Notice that IASD fluxes can be consistently included only in the presence of $\alpha'$ and non-perturbative effects which give leading contributions to the soft terms\footnote{In fact, considering IASD fluxes without including these quantum corrections, leads to phenomenological inconsistencies as found in \cite{Bena:2015qfa}.}. Finally D7 Wilson line moduli develop a scalar potential due to gauge fluxes \cite{Marchesano:2014iea}.
\een

\subsubsection{Challenges}

We shall now discuss the main challenges that type IIB models face to reach a higher level of control. Notice that most of these challenges are shared also by 4D $N=1$ supersymmetric Minkowski and AdS solutions. Hence if they are considered as indications against the existence of stable dS vacua, they should also be taken into account in criticizing existing supersymmetric solutions relevant for phenomenological applications. Here is a list of some important challenges:
\ben
\item \emph{Full control of quantum corrections}: The fact that the type IIB no-scale cancellation requires the inclusion of quantum corrections to lift the K\"ahler moduli has been the source of criticism due to the difficulty to compute all these effects in a systematic way. This criticism is indeed partially well-grounded since we are still lacking a deep understanding of both string loop corrections for arbitrary Calabi-Yau backgrounds and the exact form of $\alpha'$ corrections to the bulk and the D-brane action. Nonetheless, as discussed above, a lot of progress has been done recently, non just in computing different quantum corrections but also in estimating the volume scaling of corrections which can be neglected in the large volume limit. In this direction, as emphasised in \cite{Conlon:2006gv}, one of the most important questions is to to generalise the exact results for toroidal orientifolds in \cite{Berg:2005ja} to orientifolded Calabi-Yaus with non-zero 3-form fluxes. In fact, all other corrections relevant for the stability of the LVS vacua can be eliminated from considerations of the positivity of the K\"ahler metric \cite{Conlon:2006gv}. 

Notice that the extended no-scale structure which protects the stability of LVS models is enjoyed by any perturbative correction which is of $\mc{O}(\alpha'^2)$ regardless of the order in the string loop expansion \cite{Cicoli:2007xp}. Hence any correction to $K$ of $\mc{O}(\alpha'^2 g_s^n)$ $\forall n$ does not destabilise LVS models. Dangerous corrections would be of $\mc{O}(\alpha')$ at any order in the string coupling. However so far no contribution to the K\"ahler potential of this order has been found. In order to give a definite answer to this question it would be crucial to understand the form of string loop corrections to the K\"ahler potential in the presence of supersymmetry breaking background fluxes. 
 
Furthermore, even if in the past decade there has been substantial progress in the understanding of Euclidean D3-instantons \cite{Blumenhagen:2009qh}, we are still lacking a complete picture of these $g_s$ non-perturbative effects regarding the exact moduli-dependence of their prefactor or zero-mode lifting by gauge and background fluxes. Let us however point out that unknown $\mc{O}(1)$ coefficients of these non-perturbative effects do not tend to affect the main qualitative and quantitative results of moduli stabilisation. In addition, gaugino condensation on D7-branes has been well understood from the standard 4D effective field theory point of view but it is more difficult to study using the full 10D effective action and the full string theory.

Present technology only allows for the computation of the volumes of 4-cycles and 2-cycles after moduli stabilisation. This gives information only about the average size of the curvatures, and so in principle a 2-cycle with volume which is large in string units can be anisotropic and have regions with high curvature. While this is a challenge, it is not expected to be a generic issue. 

\item \emph{Parametrically small parameters}: No dS construction is characterised by full parametric control of the expansions used to stabilise the moduli as opposed to AdS/CFT where the $1/N$ expansion can be trusted in the large $N$ limit. While this is a fully valid theoretical objection, we argue that small expansion parameters, even if not parametrically small, are still good enough to trust the phenomenological implications of the results. A primary example is QED where the perturbative expansion is an asymptotic series which can give only an approximate result up to non-perturbative effects. One gets closer and closer to an exact result only when the expansion parameter $\alpha_\QED$ gets closer and closer to zero which in QED is the Gaussian fixed point at vanishing energy. However, experiments are performed at a fixed energy scale, and so $\alpha_\QED$ is fixed when confronting data and cannot be set arbitrarily small. Nonetheless perturbative QED yields results which reproduce data extremely well. In string compactifications, the parameter controlling the $\alpha'$ expansion is $1/\vo \ll 1$ where $\vo$ is the internal volume in string units which can be exponentially large in type IIB LVS models \cite{Balasubramanian:2005zx, Conlon:2005ki, Cicoli:2008va}. The limit $1/\vo \to 0$ would imply a string scale $M_s\sim M_{\rm p}/\sqrt{\vo}$ below the TeV scale, and so it is not phenomenologically allowed. Theoretically it is the decoupling limit of 4D gravity and leads to 10D string theory. However in our phenomenological applications we are at finite but sufficiently small values of $1/\vo$ as to be able to trust the $\alpha'$ expansion. Similar considerations apply to the string coupling which in a given point of the type IIB flux landscape is lower bounded by tadpole cancellation but can still be small enough to trust the perturbative expansion. Whether one can really justify these expansions as asymptotic series like in QED is currently unknown but hopefully may be settled at some point.

\item \emph{Supersymmetric $N=1$ vacua}: As stressed above, the main obstacle against constructing dS vacua is the difficulty to have an effective field theory which is under full control. The reason for the lack of full calculational control is the lack of supersymmetry for any dS vacuum which holds in any spacetime dimension. Therefore the apparent difficulty to obtain dS vacua may be more related to the simple fact that we do not have enough reliable techniques to tackle theories with broken supersymmetry. However it is important to stress that most of the challenges are shared also by 4D $N=1$ supersymmetric Minkowski or AdS vacua relevant for phenomenology. In fact, even if supersymmetry helps to control the structure of the effective field theory, $N=1$ models are still subject to quantum corrections whose exact form for arbitrary Calabi-Yau orientifold backgrounds has not been fully understood yet. Moreover, supersymmetric vacua do not seem very suitable starting points for phenomenological applications since they are in general characterised by flat directions which can develop a runaway behaviour when one moves away from the vacuum to break supersymmetry. A primary example of supersymmetric vacua suffering from this problem is the type IIB case where only supersymmetric $(2,1)$ $G_3$ background fluxes are turned on \cite{Dasgupta:1999ss,Giddings:2001yu}. These are $N=1$ 4D vacua where the K\"ahler moduli are flat directions. In order to be consistent with observations, one has to break supersymmetry. If this is done by moving the dilaton or the complex structure moduli away from their supersymmetric minimum, the K\"ahler moduli become unstable runaways. Alternatively, one could try to stabilise the K\"ahler moduli in a supersymmetric minimum and move them away from it to break supersymmetry while keeping the dilaton and the complex structure moduli at their minimum. However, in order to lift the K\"ahler directions, one would need to include non-perturbative effects whose 10D origin has not been fully understood yet.

\item \emph{D3-branes and sequestering}: The effective action of D3-branes at singularities is arguably the least understood aspect of 4D type IIB models. Even though a standard expansion around vanishing vacuum expectation values, as usually done for matter fields, can provide useful information \cite{Conlon:2008wa}, it would be desirable to develop a better understanding of the dependence of the K\"ahler potential on blow-up modes and open string matter fields around the singularity. A particularly interesting issue for phenomenology is a systematic study of all possible effects (perturbative and non-perturbative) which can break the sequestering of the visible sector on D3-branes from the sources of supersymmetry breaking in the bulk.

\item \emph{Explicit full moduli stabilisation}: Stabilising all closed and open string moduli in a controlled way in an explicit Calabi-Yau example is a very demanding task. Above all, because of the difficulty to solve the minimisation equations in the presence of a large number of complex structure and open string moduli. A lot of progress has been made in this direction \cite{Cicoli:2011qg, Cicoli:2012vw, Cicoli:2013mpa, Cicoli:2013cha, Cicoli:2016xae, Cicoli:2017shd, Cicoli:2017axo} (in particular models with  an effective small number of complex structure moduli) but a globally consistent model with full moduli stabilisation in a controlled dS vacuum has still to be achieved. However given the existence of a very large number of flux configurations it would be very surprising if there is no solution to these equations.

\item \emph{Realistic phenomenology}: The present accelerated expansion of our universe is just one observational feature of Nature. Other crucial characteristics of our world are chiral matter, the SM gauge group, dark matter and inflation. Hence it does not make that much sense to obtain dS vacua which cannot realise these other crucial phenomenological features. In the past few years there has been substantial progress in building global models with dS, chiral matter and inflation \cite{Cicoli:2017shd, Cicoli:2017axo} but a fully working model which can allow for both a realistic cosmology and particle physics is still missing.

\item \emph{F-theory moduli stabilisation}: Type IIB models are  the weak coupling limit of more general F-theory constructions. In order to gain more control over moduli stabilisation and D-brane model building, it is therefore fundamental to understand the F-theory uplift of the existing type IIB dS mechanisms. Another crucial issue to address is moduli stabilisation directly within the F-theory framework. 

\item\emph{Populating the landscape}: The landscape scenario to address the dark energy problem needs crucially a mechanism to populate the landscape. A concrete point is that even though the flux superpotential $W_0$ is only bounded by $W_0\ll \vo^{1/3}$, the tuning needed to address the dark energy problem requires a discretuum determined by a distribution of values of $W_0$ such that $\delta W_0$ can be made as small as possible. For this, Calabi-Yau compactifications with at least hundreds of complex structure moduli are needed to be stabilised, making the computational challenges extremely difficult. Furthermore a full quantitative understanding regarding the vacuum transitions among different solutions is not under full control yet.
\een

Having discussed the general pros and cons of IIB flux compactifications, in the following sections we will analyse more in detail the advantages and the challenges of concrete mechanisms to achieve dS vacua (see also \cite{Danielsson:2018ztv}).

\subsection{Anti-branes}
\label{Dbar}

Adding anti D3-branes to the KKLT and LVS setups provides a simple positive contribution to the vacuum energy coming from their tension. This is the concrete KKLT proposal based on the KPV construction \cite{Kachru:2002gs} on brane-flux annihilation for which an anti-brane sitting at the tip of a throat induced by the 3-form fluxes can annihilate with the fluxes after polarising a NS5-brane which later decays. This process can be described in terms of quantum tunneling through a barrier. Fluxes can be tuned to control the size of the throat that can be used to adjust the vacuum energy to desired values. As emphasised above, this mechanism can be considered together with fluxes and non-perturbative effects although in the original presentation it was introduced as an uplift mechanism of the original AdS vacuum providing a positive correction to the scalar potential of the form:
\be
\Delta V = \frac{e^A}{\vo^\gamma}\,,
\label{uplift}
\ee
with $e^A$ the flux-induced warp factor and $\gamma=4/3$ in the warped region while $\gamma=2$ in an unwarped region. The warp factor can be used to tune the minimum to dS at almost zero vacuum energy.

\subsubsection*{Criticism 1}

Since it was proposed, this has been considered as the weakest part of the KKLT proposal. Despite the relation with the KPV scenario, adding an anti-brane seems  arbitrary. It also seems to break supersymmetry explicitly, losing computational control of the EFT and giving a runaway behaviour to 10D at the classical level\footnote{For an early discussion of the problems with anti-branes in KKLT see \cite{Brustein:2004xn}.}. Furthermore the original scenario was not substantiated by explicit models on concrete Calabi-Yau orientifolds. More recently detailed study of the geometry corresponding to anti-branes on a throat indicated the presence of singularities that were claimed to destabilise the KKLT system if anti-branes were present \cite{Bena:2009xk, Bena:2012bk}.

\subsubsection*{Comments}

The anti-brane sector has been probably the most questioned component of the KKLT proposal. Regarding the apparent arbitrariness, the KPV scenario already provides a natural motivation for its consideration. The fact that supersymmetry is broken has been better understood by the recent developments relating the EFT of the anti-brane to non-linearly realised supersymmetry {\it a la}  Volkov and Akulov. Moreover, a concrete superspace formulation in terms of a nilpotent chiral superfield $X$ ($X^2=0$) \cite{Ferrara:2014kva, Bergshoeff:2015jxa} captures precisely the term in \eqref{uplift} by adding to the original superpotential and K\"ahler potential  a general dependence on $X$:
\be
\Delta W= c \,X\,,  \qquad \Delta K = \beta\, X\bar{X}\,.
\ee
Here $c$ is in principle a function of the complex structure moduli which can be naturally associated to warping while $\beta$ depends also on the K\"ahler  moduli. The superfield $X$ has a single propagating degree of freedom corresponding to the goldstino. Concrete Calabi-Yau orientifolds have been constructed (compact and non-compact) with precisely this single degree of freedom \cite{Kallosh:2015nia, Garcia-Etxebarria:2015lif}, so providing the first explicit realisations of the dS KKLT scenario. Finally an EFT analysis of the anti-brane singularity has been done for the simplest case of one single anti-brane (which is sufficient to achieve dS) for which the probe approximation is under control and no divergences are found, so addressing the anti-brane induced singularity problem \cite{Michel:2014lva, Polchinski:2015bea}. The same conclusion has been reached recently using different techniques \cite{Cohen-Maldonado:2015ssa, Cohen-Maldonado:2015lyb, Danielsson:2018ztv}.

\subsubsection*{Criticism 2}

Another potential obstacle has been claimed by ref. \cite{Moritz:2017xto} regarding the calculation of non-perturbative effects when anti-branes are present. In an effort to have a 10D description of gaugino condensation, ref. \cite{Moritz:2017xto} developed a technique to compute the contribution of the anti-brane to the scalar potential and found no dS solution. This was understood also from the 4D EFT in terms of the nilpotent superfield $X$ by considering the $X$ dependence of $W$ as:
\be
\Delta W= X\left(c+e \,W_{\rm np}\right).
\ee
It is easy to check that for $c=0$ and $e\neq 0$ the contribution of $X$ to the scalar potential is such that there is no dS vacuum either in KKLT \cite{Moritz:2017xto} or in LVS \cite{MQRW}.

\subsubsection*{Comments}

The result regarding the non-perturbative superpotential in the presence of anti-branes is based on a number of different assumptions which are not fully justified. The most relevant is perhaps assuming that the dynamics of gaugino condensation $\langle \lambda \lambda\rangle $ can be described in terms of the $\lambda\lambda$ dependence of the classical action. Gaugino condensation is clearly a 4D non-perturbative effect due to the non-trivial low-energy dynamics of the corresponding gauge theory. Its effect needs to be computed by properly performing the path integral of the gauge degrees of freedom below the scale of the relevant gauge theory which is a highly complicated quantum calculation. It is actually known in field theory that properly computing the effective superpotential does not reproduce the result of naively substituting $\lambda\lambda\sim \Lambda\strong$ in the classical effective action where $\Lambda_\strong$ is the condensation scale. In fact, at least in the case of the heterotic string, one can show the conflict quite explicitly \cite{Brustein:2004xn,Nilles:2004zg}.

From the 4D EFT perspective, the fact that the coefficient $c$ vanishes in $\Delta W$ (needed to avoid dS) does not seem justified, especially since this leads to the term $X W_{\rm np}$ to be the dominant contribution which is instead expected to be very much suppressed since, besides the non-perturbative suppression, the coefficient $e$ is naively expected to be suppressed by warp factors (as for $c$ in the original case). Also in the absence of non-perturbative effects it is known that $c\neq 0$. Furthermore the analysis in \cite{Moritz:2017xto} does not include the case of Euclidean D3-instanton contributions to the scalar potential. The proposal presented in \cite{Moritz:2017xto} to obtain dS by considering a racetrack scenario, while possible, has not been implemented in concrete models and may be difficult to construct without fine-tuning coefficients of the non-perturbative terms.

\subsection{T-branes}
\label{T}

T-branes represent a very generic and natural way to obtain dS vacua in type IIB models via the interplay of background and gauge fluxes. From the 4D point of view, this mechanism relies on non-zero F-terms of hidden sector fields driven by D-term stabilisation. The 8D understanding of this system involves a background with a T-brane which is a non-Abelian bound state of D7-branes induced by gauge fluxes. Expanding the T-brane action in the presence of supersymmetry breaking background fluxes gives rise to a positive definite contribution to the 4D scalar potential which can be used to obtain a dS vacuum \cite{Cicoli:2015ylx}. Let us stress that this approach to dS vacua has been used in several explicit global Calabi-Yau models \cite{Cicoli:2012vw, Cicoli:2013mpa, Cicoli:2013cha, Cicoli:2017shd}.

The generality of this mechanism is based on the following observations:
\bi
\item D7 tadpole cancellation in models with O7-planes forces in general the presence of hidden sector stacks of D7-branes.

\item The absence of Freed-Witten anomalies on D7-branes requires in general to turn on half-integer gauge fluxes on the worldvolume of D7-branes \cite{Minasian:1997mm, Freed:1999vc}.

\item These gauge fluxes induce K\"ahler moduli-dependent Fayet-Iliopoulos terms \cite{Dine:1987xk,Dine:1987gj} which are cancelled by a non-zero vacuum expectation value of an open string mode charged under the corresponding anomalous $U(1)$. Hence this Abelian gauge group is broken and the corresponding gauge boson acquires a mass of order the string scale. This signals the fact that the anomalous $U(1)$ should not have been included in the 4D effective field theory. In fact, the correct low-energy theory should be built by expanding around a non-zero vacuum expectation value of the open string mode (as opposed to a vanishing value which would seem to lead to a $U(1)$ factor) which corresponds to a T-brane background characterised by a non-Abelian bound state of D7-branes without any $U(1)$ factor.

\item The tree-level GKP solution features $(0,3)$ background 3-form fluxes which break supersymmetry. 

\item These supersymmetry breaking fluxes induce soft term masses for open string modes on D7-branes which correspond to F-term contributions to the scalar potential \cite{Camara:2004jj}.

\item Expressing both the soft term masses and the vacuum expectation value of the charged open string mode in terms of the K\"ahler moduli, one obtains a positive definite volume-dependent contribution to the scalar potential which can raise the vacuum energy to dS.
\ei
Let us point out that, contrary to what has been claimed in \cite{Bena:2017uuz}, the dS construction of \cite{Cicoli:2015ylx} does not require T-branes in the strong coupling regime since gauge flux densities are always below the string scale. In fact, the 8D BPS equation determining the T-brane background receives perturbative corrections which cannot be neglected when the Higgs vacuum expectation value is above the string scale, implying a large gauge flux density \cite{Minasian:2001na, Marchesano:2016cqg}. In this strong coupling regime, at least for the $N=2$ case with a large number of branes, the system has been shown to be describable in a dual picture which involves a single Abelian D-brane without flux but with worldvolume curvature \cite{Bena:2016oqr}. However the T-brane setup of \cite{Cicoli:2015ylx} is in the weak coupling regime where the effective field theory is under control since the volume in string units of the 2-cycle supporting the gauge flux is always much larger than the flux quanta. The corresponding 4D Higgs field develops a vacuum expectation value of order the winding scale $M_\W \sim M_s \vo^{1/6} \sim M_{\rm p} /\vo^{1/3}$. Taking into account the right 4D field in Einstein frame, the condition of negligible perturbative corrections to the 8D BPS equation is a 4D Higgs field below the scale $M \sim M_{\rm p} / \vo^{1/6}$. For $\vo \gg 1$ the winding scale is below $M$, and so $\alpha'$ corrections to the Fayet-Iliopoulos term can be safely ignored. 

Let us finally mention that this dS mechanism has two limitations: 
\ben
\item The D-term is proportional to the sum of the F-terms of the volume modulus and the charged open string mode. In KKLT models the F-term of the $T$-moduli is vanishing, and so a vanishing D-term necessarily implies also that the F-term of the Higgs field has to be zero. Hence dS KKLT vacua cannot be obtained via T-branes. On the other hand, in LVS models the F-term of the charged K\"ahler modulus is non-zero. This guarantees that the D-term can be set to zero (at least at leading order) in a way compatible with a non-vanishing F-term of the charged open string mode which determines the T-brane background. 

\item The volume-dependence of the positive definite contribution to the scalar potential from T-branes is $0.01 c \,\vo^{-8/3}$ where $c$ depends on gauge flux quanta. On the other hand, the LVS potential generated by $\alpha'$ and non-perturbative effects scales as $\vo^{-3} \sqrt{\ln\left(\frac{\vo}{W_0}\right)}$ \cite{Cicoli:2015ylx}. Given that the flux quanta are $\mc{O}(1)$ integer parameters, the only quantity which can be tuned to obtain a dS vacuum is the tree-level superpotential $W_0$. For natural $\mc{O}(1-10)$ values of $W_0$, T-branes allow for dS vacua with values of the volume in string units of order $\vo\sim 10^5-10^8$. Interestingly this is in the right ballpark to get low-energy supersymmetry in sequestered scenarios with the MSSM on D3-branes at singularities \cite{Aparicio:2014wxa, Blumenhagen:2009gk}. However if the volume is raised to values of order $\vo\sim 10^{15}$, as needed to obtain TeV-scale soft terms in non-sequestered models \cite{Conlon:2005ki}, the T-brane contribution to the scalar potential would yield a runaway for the volume mode. Notice that this instability cannot be avoided by tuning $W_0$ extremely small since the vacuum expectation value of the volume mode is also proportional to $W_0$. This implies that in T-brane dS vacua with non-sequestered visible sector, the scale of supersymmetry breaking has necessarily to be rather high. 
\een

\subsection{Other dS mechanisms}
\label{OtherdS}

Since the original KKLT model several other mechanisms have been proposed mostly using string inspired field theoretical arguments in which F and/or D terms yield positive contributions to achieve dS (see for instance \cite{Burgess:2003ic, Villadoro:2005yq, Dudas:2006gr}). Even though these are not concrete string theory models, for most of these scenarios there may be a way that some of them can eventually find a stringy realisation. The main point is that there is a wide diversity of model dependence on the matter sector of the hidden sectors. Since the scalar potential of supergravity is not positive definite it is very plausible that there exist minima with all signs of the vacuum energy. The challenge is to have concrete compactifications in which the proposed F and D terms are realised with the matter content, matter superpotential and K\"ahler potential under control and that the corresponding extrema lie in regions of the moduli space in which the EFT can be trusted. 

We can highlight eight other string motivated proposals:
\begin{itemize}

\item {\it Non-critical strings} \cite{Maloney:2002rr}: Non-critical strings have a natural positive cosmological term (for $D>10$) that can be used to obtain dS upon compactification while fixing the moduli. It is not clear whether the corresponding EFT is  under control but this may be due only to our current limited understanding of the theory.

\item {\it Negative curvature spaces} \cite{Silverstein:2007ac}: Non-supersymmetric compactifications on manifolds with negative curvature naturally induce a positive term in the effective potential that can be used to obtain dS. Again, being non-supersymmetric, the EFT is under less control but these compactifications are in principle viable.

\item {\it K\"ahler uplift} \cite{Westphal:2006tn, deAlwis:2011dp, Rummel:2011cd}: Here $\alpha'$ corrections to the K\"ahler potential in the KKLT scenario can compete with the fluxes and the non-perturbatively effects to produce minima with positive vacuum energy. This is possible if the volume is small enough for $\alpha'$ corrections to be relevant. Therefore the obtained dS minima are in regions at the edge of validity of the EFT. An explicit CY compactification has been constructed in \cite{Louis:2012nb} with all geometric moduli stabilised to dS space.

\item {\it Dilaton dependent non-perturbative effects} \cite{Cicoli:2012fh}: In type IIB models hidden and observable sectors can be localised on either D3 or D7-branes. In general, non-perturbative effects depend on the $T$-modulus which controls the volume of the divisor wrapped by an ED3-instanton or by a D7-stack supporting gaugino condensation. Another possibility is however to consider dilaton-dependent non-perturbative effects coming from E($-1$)-instantons or strong dynamics on a hidden sector of D3-branes at singularities. In this case, after including the shift of the dilaton proportional to the blow-up mode resolving the singularity, the corresponding non-perturbative term generates a positive definite contribution to the scalar potential similar to that coming from anti-branes. Concrete CY compactifications with these superpotentials have not been constructed yet and it would be an interesting avenue to explore.

\item {\it Complex structure F-terms} \cite{Gallego:2017dvd}: The complex structure moduli have minima at the supersymmetric points $D_UW=0$. However there may be further minima for these fields for which $D_UW\neq 0$. These may give rise to dS minima without the need of further ingredients but need to tune quantities such that the corresponding new minimum, coming from $\mc{O}(1/\vo^2)$ terms in the scalar potential, does not wash out the large volume minimum coming from terms of order $\mc{O}(1/\vo^3)$. An intrinsic limitation of these constructions is the difficulty to realise explicit examples with large volume since they would require to fix a large number of complex structure moduli to achieve enough tuning freedom. A concrete example with $\vo\simeq 10^4$ was however constructed in \cite{Gallego:2017dvd} but using a continuous flux approximation.

\item {\it Non-perturbative dS vacua} \cite{Blaback:2013qza}: dS minima
can emerge from stabilising all the geometric moduli in just one-step via
the inclusion of just background fluxes and non-perturbative effects. The
main problems of this approach are the poor knowledge of the $S$ and
$U$-moduli dependence of the prefactor of non-perturbative effects and the
computational difficulty to find a numerical solution for the minimisation
equations in the presence of a large number of geometric moduli.

\item \textit{Heterotic dS vacua} \cite{Cicoli:2013rwa}: The heterotic string compactified on smooth Calabi-Yau threefolds can lead to 4D dS models where the gauge bundle moduli, together with the dilaton and the $U$-moduli are fixed supersymmetrically at leading order via the requirement of a holomorphic gauge bundle, fractional fluxes and non-perturbative effects. The K\"ahler moduli can instead be fixed a la LVS by the interplay of worldsheet instantons, $\alpha'$ effects and threshold corrections to the gauge kinetic function which provide a positive term responsible for achieving a dS vacuum. The main limitations of this construction are: $(i)$ the phenomenological value of the GUT gauge coupling forces the minimum to lie only at moderately large volume and $(ii)$ the lack of tunability of the vacuum energy due to the absence of Ramond-Ramond fluxes.

\item {\it $G_2$ compactifications} \cite{Acharya:2006ia,Acharya:2007rc}: Even though little is known about concrete $G_2$ holonomy compactifications of M-theory, interesting scenarios have been proposed addressing phenomenological issues. In particular, superpotentials with two exponentials but without fluxes have been proposed to stabilise all moduli. Some of the minima can be dS even if the absence of fluxes makes it difficult to tune the vacuum energy to small values. The freedom to perform this tuning relies on the possibility to find a small value of the cosmological constant by scanning through different ranks of the condensing gauge groups.
\ei

\section{Quintessence in string theory}
\label{sec:QuintessenceStrings}

In the previous section we have argued that there are explicit string theoretic constructions which seem to violate the conjecture (\ref{eq:SwamplandConjecture}) even if there are still some technical issues that have to be fully sorted out. In this section we will argue that the alternative, i.e. potentials which satisfy the conjecture and lead to quintessence models, are far less likely to be consequences of string theory.

The simplest alternative to a vacuum energy of order $\langle V\rangle_0 \sim \Lambda^4 \sim ({\rm meV})^4$ is a scalar field which is slow-rolling at positive energies. The dynamics of this scalar field is driving the present epoch of accelerated expansion (for a review see \cite{Tsujikawa:2013fta}). A simple example of such a phenomenological potential would be\footnote{In this section we work in Planck units except where other units are explicitly invoked for clarity.} $V= \Lambda^4\,e^{-\chi}$ where $\Lambda$ is the current dark energy scale at $\chi=0$ today. 

In the presence of a plethora of scalar fields like in the string landscape, it might seem rather natural to expect that one of them is at present rolling away from its minimum. This picture seems also to be suggested by the recently proposed swampland conjecture which forbids the existence of stable dS solutions \cite{Agrawal:2018own}. Even if this is still a conjecture not based on any rigorous derivation, it is worth exploring its phenomenological implications. In what follows we shall discuss in particular the interplay of quintessence and the swampland conjecture with the Higgs potential and low-redshift cosmological data.

\subsection{Challenges for quintessence}
\label{sec:QuintessenceChallenges}

The main challenges for quintessence are:
\bi
\item \textit{Fine-tuning problems}: Standard dS models are plagued by the problem of obtaining the right value of the cosmological constant. Quintessence models share the same problem together with an additional fine-tuning problem related to the necessity to obtain a scalar field which is light enough to drive the present epoch of accelerated expansion. In fact, why should the quintessence field be today exactly at the point where $V\simeq \Lambda^4$? Moreover, in order to have a working model, the quintessence field has to be extremely light with $m\simeq 10^{-32}$ eV. How can one make a scalar field so light? What symmetry is protecting the mass of this field?

In fact, the quintessence potential is usually given as a phenomenological construct valid at cosmological scales. In this case any quantum corrections
are also of the same order (i.e. $\mc{O}(\Lambda^4)$ since the cutoff is of order $\Lambda$), and so will not destabilise this potential. However if one is to obtain this from some more fundamental theory at a higher scale, then this immediately faces a problem of destabilisation. This is the old problem of the cosmological constant but with an additional twist since we need to preserve not only the vacuum energy but also the running of the latter with the rolling of the quintessence field\footnote{For a detailed discussion of quantum corrections see \cite{Garny:2006wc}.}.

For instance consider a simple quintessence model with potential $V= \Lambda^4\,e^{-\chi} + m^2 \phi^2$ where $\chi$ is rolling while $\phi$ is a massive field. After integrating out the field $\phi$, which is sitting at the minimum of its potential, the quintessence model at the cutoff scale looks like:
\be
V=\Lambda^4\, e^{- \chi}+\frac{1}{32\pi^2}\,m_\phi^4+\ldots\,.
\ee
where the ellipses indicate lower order terms. In the presence of fermions, there will be corresponding negative contributions to the right-hand side but unless there is unbroken supersymmetry at these scales there will be a positive contribution that is much larger than the current cosmological constant. If one now tunes the vacuum energy (choosing $\chi=0$ at present), then to have the current value of dark energy we need: 
\be
\Lambda^4 =-\frac{1}{32\pi^2}\,m_{\phi}^4-\ldots\,.
\ee
Thus this will destabilise the potential. In general in such a model to restore the phenomenological quintessence model one would need to do functional fine-tuning. It is also possible that there are quintessence models in which the potential takes the form: 
\be
V=\Lambda^4\, e^{- \chi}+V_0\,.
\ee
In which case the quantum fluctuations of $\phi$ can be absorbed by fine tuning $V_0$. In this case we would be left just with two fine-tunings (as in generic quintessence models) i.e. one more than in generic dS models. However, in the absence of an underlying symmetry protecting the quintessence potential, it is not clear to us whether such a model (i.e. with $V_0\ne 0$ and fine tunable) can be constructed that will satisfy the conjecture of \cite{Obied:2018sgi}.

Moreover, let us point out that in a string inspired supergravity setup, after supersymmetry breaking, the corrections to the quintessence field potential would be of order $m_{3/2}^2 \Lambda_\cutoff^2$. Given that the gravitino mass $m_{3/2}$ sets the scale of the soft-terms, $M_{\rm soft} \sim m_{3/2}$, it cannot be smaller than the TeV scale. Since the cutoff scale $\Lambda_\cutoff$, which in 4D string models is naturally given by the Kaluza-Klein scale, has to be larger than $m_{3/2}$ in order to control the effective field theory, we conclude that the quintessence potential will generically receive corrections which cannot be smaller than (TeV)$^4$. This poses a serious challenge for any quintessence model.

\item \textit{Phenomenological problems}: The quintessence field can be either a scalar or a pseudo-scalar. If it is a pseudo-scalar like an axion, it can avoid fifth-force constraints, but a typical axion potential is flat enough to drive a period of accelerated expansion only if its decay constant is trans-Planckian \cite{Freese:1990rb}. This is in disagreement with recent studies of axion field ranges from string theory \cite{Klaewer:2016kiy, Blumenhagen:2017cxt, Palti:2017elp, Cicoli:2018tcq, Grimm:2018ohb, Heidenreich:2018kpg}. On the other hand, if the quintessence field is a scalar, it is not clear how to avoid the existing stringent bounds from fifth-forces \cite{Adelberger:2003zx}. Moreover, if the quintessence field is a string modulus which sets the visible sector gauge kinetic function, a rolling modulus would give rise to a time variation of the coupling constants. This last problem can be avoided simply by considering a modulus which is not supporting the visible sector stack of D-branes. However, evading fifth-force bounds is more complicated. The volume mode couples democratically to all fields with Planckian strength, and so it cannot be the quintessence field. This is a direct consequence of the locality of the SM construction. The fact that the volume mode has to couple to SM fields can be seen by looking at the relation between the physical Yukawa couplings $\hat{Y}_{ijk}$ and the holomorphic ones $Y_{ijk}(U)$ which depend just on the complex structure moduli because of the holomorphicity of the superpotential and the axionic shift symmetry \cite{Conlon:2006tj}:
\be
\hat{Y}_{ijk} = e^{K/2}\,\frac{Y_{ijk}(U)}{\sqrt{\tilde{K}_i \tilde{K}_j \tilde{K}_k }}\,,
\label{Yuk}
\ee
where $\tilde{K}_i$ is the K\"ahler metric for matter fields. Due to locality, the physical Yukawa couplings should not depend on the overall volume, and so the matter K\"ahler metric $\tilde{K}_i$ has to depend on the volume mode $\vo$ in order to cancel the powers of $\vo$ in $e^{K/2}$. Consequently, the volume mode has always a direct $M_{\rm p}$-suppressed coupling to SM-fields from expanding the matter K\"ahler metric in the kinetic terms. 

The best case scenario is therefore when the quintessence field is a modulus different from the overall volume which supports a hidden sector stack of branes, while the visible sector is localised on a blow-up mode which does not intersect with the quintessence divisor. This has been advocated in the context of swampland conjectures in \cite{Agrawal:2018own}. However even in this case, one would need to check that no interaction between the quintessence modulus and visible sector fields is induced by kinetic mixing between the moduli (see for example the moduli redefinitions in \cite{Conlon:2007gk,Cicoli:2010ha, Cicoli:2012cy} induced by non-canonical kinetic terms) or between hidden and visible sector Abelian gauge bosons \cite{Abel:2008ai, Burgess:2008ri, Goodsell:2009xc, Cicoli:2011yh}. This issue is currently under detailed investigation \cite{AMM}. 
\ei

\subsection{The swampland and the Higgs}
\label{sec:SwampandHiggs}

As already pointed out in \cite{Denef:2018etk}, the swampland conjecture is in tension with basic features of the Higgs potential. In fact if $h$ is the standard Higgs field and $\chi$ the quintessence field, the total scalar potential can be written as:
\be
V = \tilde{V}(h) + \hat{V} (\chi)\qquad\text{with}\qquad \tilde{V}(h) = \lambda \left(h^2-v^2\right)^2 \,.
\label{Vin}
\ee
The swampland conjecture at the maximum of the Higgs potential for $h=0$ then implies:
\be
\frac{|\nabla V|}{V}\gtrsim 1\qquad \Leftrightarrow\qquad \frac{\hat{V}_\chi(\chi)}{\tilde{V}(h) + \hat{V} (\chi)} = \frac{\hat{V}_\chi(\chi)}{\lambda v^4 + \hat{V} (\chi)} \gtrsim 1\,.
\label{NoSwamp}
\ee
However the quintessence potential today has to scale as $\hat{V}(\chi_0)=\Lambda^4$. Typical quintessence potentials have the form $\hat{V}(\chi)=\Lambda^4\,e^{-\chi}$ with $\chi_0\simeq 0$. Hence $\hat{V}_\chi(\chi_0)\simeq \hat{V}(\chi_0)=\Lambda^4$, implying that the ratio in (\ref{NoSwamp}) violates the swampland conjecture by $57$ orders of magnitude!

There are several ways to cure this problem but none of them seems very natural from the string theory point of view:
\bi
\item \textbf{Higgs as quintessence:} As a first pass at a solution one might ask whether the quintessence field can be identified with the Higgs field itself along the lines for instance of Higgs inflation, modified appropriately for quintessence. In this case at low energies (below the scale of electroweak breaking) the Higgs potential (for the neutral Higgs in unitary gauge) may acquire the form:
\be
V = \Lambda^4 + C^4 \,e^{- k\,h/M_{\rm p}}\,.
\ee
Imposing that the Higgs is rolling today at $h=v$ with values of the slow-roll parameter $\epsilon = \frac{M_{\rm p}^2}{2} \left(\frac{V_h}{V}\right)^2$ of order $1/2$ and $V \simeq \Lambda^4$ together with the right Higgs mass, one can fix the values of the parameters $C$ and $k$ at $C \simeq 10^{-52} e^{2.5\cdot 10^{71}} M_{\rm p}$ and $k=10^{88}$. Notice that this model is in agreement with observational data since, due to the huge value of $k$, one can get around $5$ efoldings of exponential expansion for $\Delta h\simeq 10^{-85.7}\,M_{\rm p}$, implying that no time-variation of the fermion masses could be observable. However the unreasonable value of $k$ and $C$ show that this is more a curious observation rather than a real solution.

\item \textbf{A direct Higgs-quintessence coupling:} Ref. \cite{Denef:2018etk} modified the initial potential (\ref{Vin}) via the introduction of a coupling between $\chi$ and $h$ of the form:
\be
V = f(\chi) \,\tilde{V}(h) + \hat{V} (\chi)\qquad\text{with}\qquad f(\chi) = e^{-\chi} \,.
\label{Vnew}
\ee
In this case the swampland conjecture is satisfied since the ratio in (\ref{NoSwamp}) at $h=0$ where $\tilde{V}_h (h)=0$ takes the form:
\be
\frac{f_\chi(\chi) \,\tilde{V}(h)+\hat{V}_\chi(\chi)}{f(\chi)\,\tilde{V}(h) + \hat{V} (\chi)} \simeq  \frac{f_\chi(\chi)}{f(\chi)} \simeq 1\,.
\label{YesSwamp}
\ee
However, even if the Higgs-quintessence coupling in (\ref{Vnew}) is not ruled out by fifth-force constraints \cite{Denef:2018etk}, one would need to explain why the SM fermions are instead decoupled from the quintessence field since a direct coupling between them and $\chi$ would not be allowed by fifth-force bounds. Given that in 4D string models a direct coupling between $\chi$ and $h$ would generically also imply a direct coupling between the quintessence field and SM fermions, we interpret this tension as a phenomenological hint against the validity of the swampland conjecture. 

\item \textbf{Adding more fields}: Another solution involves the introduction of a third field $\phi$ which is heavy in the electroweak vacuum but makes a non-trivial contribution to the criterion at the symmetric point of the Higgs potential. Hence the potential (\ref{Vin}) gets modified to:
\be
V = f(\phi)\,\tilde{V}(h) + g(\phi) + \hat{V} (\chi)\,.
\label{Vin2}
\ee
Defining the function $y(\phi)\equiv \lambda v^4\,f(\phi) + g(\phi)$, the swampland criterion (\ref{NoSwamp}) evaluated at $h=0$ then looks like:
\be
\frac{\sqrt{y_\phi^2(\phi)+\hat{V}_\chi^2(\chi)}}{y(\phi) + \hat{V} (\chi)} \simeq  \frac{y_\phi(\phi)}{y (\phi)} \gtrsim 1\,.
\label{NoSwamp2}
\ee
Notice that $y_\phi(\phi)$ corresponds the gradient of the potential in the $\phi$ direction at the symmetric point of the Higgs potential. Thus the field $\phi$ can help to satisfy the swampland criterion if $y_\phi(\phi)\simeq y (\phi)$.

On the other hand, the same ratio evaluated at the minimum of the Higgs potential at $h=v$ becomes:
\be
\frac{\sqrt{g_\phi^2(\phi)+\hat{V}_\chi^2(\chi)}}{g(\phi) + \hat{V} (\chi)} \simeq  \frac{g_\phi(\phi)}{g (\phi)} \gtrsim 1\,.
\label{NoSwamp3}
\ee
In order for $\phi$ to be stabilised at the present Higgs vacuum at $h=v$, we need also to impose that the function $g (\phi)$ admits at point in field space $\phi_0$ where $g_\phi(\phi_0) = g (\phi_0) = 0$ so that (\ref{NoSwamp}) is satisfied. Moreover, we need also to require that $f(\phi_0)=1$ in order to obtain a massive Higgs field. 

We now turn to writing down explicit models where the criterion is satisfied and point out the challenges in realising them:

\underline{Model 1:}

The simplest choices for the functions $y(\phi)$ and $g(\phi)$ satisfying the requirements (\ref{NoSwamp2}) and (\ref{NoSwamp3}) are $y(\phi)=\lambda v^4\,e^\phi$ and $g(\phi)= m^2 \Delta\phi^2$ with $\Delta\phi\equiv\phi-\phi_0$, which imply a coupling function of the form $f(\phi) = e^{\phi} - \frac{m^2}{\lambda v^4}\,\Delta\phi^2$. Notice that the potential (\ref{Vin2}) admits a point with $h$ different from either zero or $v$ where the swampland conjecture might seem to be violated since $V_\phi=V_h=0$ with positive energy. In fact, $V_h=0$ can be solved also for a negative value of $\Delta\phi$ such that $e^{\phi_0} e^{\Delta\phi} = \frac{m^2}{\lambda v^4}\,\Delta\phi^2$, and then a certain value of $h$ would solve $V_\phi=0$. However, if $\phi_0\gg 1$ and $m$ is not too large, this solution would require $|\Delta\phi|>1$, in a regime not allowed by the swampland conjecture on field distances. Notice that if the microscopic origin of the auxiliary field $\phi$ is a string modulus, we expect $\phi_0\gg 1$ in order to trust the effective field theory. As an example, consider the volume mode $\tau$ which needs to be fixed at values much larger than unity and would be related to the canonically normalised field $\phi$ by the transformation $\phi = \sqrt{\frac32} \ln\tau$. The upper bound on $m$ can be obtained by setting $\vo=\tau^{3/2}\lesssim 10^{30}$ which would correspond to TeV-scale strings. In turn, $\phi_0\lesssim 56$ which would require $m\lesssim 1$ MeV to have the extra solution discussed above at $|\Delta\phi|>1$. If we instead set $\vo\lesssim 10^{15}$ which would correspond to a TeV-scale gravitino, $\phi_0\lesssim 28$ which would require $m\lesssim 1$ eV. These values of $m$ are still large enough to be able to couple $\phi$ to both the Higgs and any SM fermion without introducing any fifth-force. Hence this solution does not feature any observational problem but does not look very appealing since it requires a particular form of the coupling function $f(\phi)$. Notice, in particular, that an order one change in the coefficient of the $\phi$-Higgs coupling proportional to $m^2$ would violate the conjecture.

\underline{Model 2:}

One could consider the more standard case where $f(\phi) =  e^{\phi}$ which implies Planck suppressed couplings between the Higgs and $\phi$ whose field range is taken to be $|\phi| < 1$ (motivated by the swampland conjecture on field ranges) with $\phi=0$ at the present vacuum. The requirement (\ref{NoSwamp2}) can still be satisfied with $y(\phi)\simeq \lambda v^4\,e^\phi$ if the function $g(\phi)$ is such that it is always subleading with respect to $\lambda v^4\,e^\phi$ for $|\phi| < 1$. This is true if $g(\phi)$ has a bounded range with a maximum and a minimum well below the electroweak scale. A simple way to realise this is if $g(\phi)$ tends to a constant value, i.e $g(\phi)$ is linear in regions away from its minimum. To analyse the case of the linear potential, we take $g(\phi) = \mu^2 \tilde{\Lambda}^2 h (\phi/ \tilde{\Lambda})$ where $h(x)$ satisfies $h(0) = h'(0) = 0$, $h''(0) = \mc{O}(1)$, and $h(x) \sim x$ for $x \gg 1$\footnote{A function with the listed properties is $h(x) = \ln( \cosh(x))$.}. With this, the mass of $\phi$ at the $\phi=0$ minimum is of order $\mu$. The parameter $\tilde{\Lambda}$ appears in the Taylor expansion about the origin in field space. The condition (\ref{NoSwamp2}) is then satisfied if $\mu^2 \tilde{\Lambda} \ll \lambda v^4$. Taking $\tilde{\Lambda}$ to be order of $\mu$ this implies $\mu < 0.5 \,{\rm MeV} \simeq m_e$. Notice that this mass is well within the fifth force bounds if $\phi$ interacts with the entire SM sector with Planck suppressed couplings\footnote{A similar mechanism can also be implemented for the low energy QCD effective potential.}. However if in the early universe the field is displaced from its minimum, it would decay after BBN creating cosmological problems. In Appendix A we check the validity of the conjecture for all points in field space. This solution features a standard coupling function $f(\phi)=e^\phi$ but it requires $g(\phi)$ to have a particular property, i.e $g_{\phi}(\phi)$ has a bounded range with a maximum and a minimum well below the electroweak scale. This will generically imply that that the self interactions of $\phi$ are set by the very low scale $\mu$. Furthermore, from the scaling dimensions $\phi$ has the properties of a fundamental scalar and not a composite. One possibility is that the field $\phi$ is localised in a warped region. One could also think of alleviating this tension by considering potentials $g(\phi)$ in which the mass is not correlated with the asymptotic value of the energy, but this seems unnatural.

\underline{Model 3:}

Another possibility to satisfy (\ref{NoSwamp2}) is the case where the gradient of the potential along the $\phi$ direction can vanish at a point in field space if also $y(\phi)=0$ at the same point. This would correspond to the case where $y(\phi) = m^2(\phi-\mu)^2$ and $g(\phi) = m^2 \phi^2$ with $m = \sqrt{\lambda} v^2/\mu$ which implies a coupling function of the form $f(\phi) = 1-2 \frac{\phi}{\mu}$. For $\mu=M_{\rm p}$, one would recover a standard modulus with Planck suppressed interactions to matter fields. However in this case the mass of $\phi$ would need to be $m = \sqrt{\lambda} v^2/ M_{\rm p} \simeq 0.01$ meV which is too low to satisfy present fifth-force bounds \cite{Adelberger:2003zx}. Hence the scale $\mu$ has to be smaller than $M_{\rm p}$. Moreover, we stress again that $\mc{O}(1)$ changes in the coefficients of $f(\phi)$ would invalidate this solution.
\ei
Notice that a way to address this challenge between the Higgs potential and the swampland conjecture is to modify the conjecture to allow saddle points at positive values of the potential as recently suggested in \cite{Andriot:2018wzk}. This could also take care of the recent observation regarding the existence of dS critical points associated with any supersymmetric KKLT-like AdS vacuum \cite{Conlon:2018eyr}. 

\subsubsection*{Supergravity issues}

In the previous section we discussed how to avoid a coupling between the Higgs and the quintessence field within the context of a non-supersymmetric theory. However if the 4D effective action coming from string theory is an $N=1$ supergravity  it seems hard to get the decoupled structure that we have suggested\footnote{See \cite{Denef:2018etk} for an earlier discussion on some challenges for quintessence in the context of supergravity. Our treatment is more general.}. This is a simple consequence of the fact that the potential has to be written in terms of a K\"ahler potential and a superpotential if one assumes that, after all the heavy moduli are integrated out, one still has the field content of a supergravity. We shall now consider a particular supergravity setup which seems to be the best starting point to realise a quintessence field decoupled from the Higgs, but we shall in the end show that the intrinsic nature of supergravity always induces an $M_{\rm p}$-suppressed coupling between the two fields.

In $N=1$ supergravity, the potential has the form (setting $M_{\rm p} = 1$ and ignoring D-terms which would not change our conclusion):
\be
V(\chi,\Phi)=e^{K(\chi,\Phi)}(|D_{\chi}W|^{2}+|D_{\Phi}W|^{2}-3|W|^2) \,.
\label{eq:V}
\ee
where $\chi$ is the quintessence field and $\Phi$ collectively denotes matter fields\footnote{Notice that this example does not contain the case of \cite{Cicoli:2012tz} which we shall review in Sec. \ref{sec:StringQuintessenceModels}. In this model the quintessence field $\chi$, being a K\"ahler modulus orthogonal to the volume mode, does not appear in the tree-level K\"ahler potential. This is a key feature to guarantee the absence of $M_{\rm p}$-suppressed couplings of $\chi$ to SM fields.}. In this context the most one can do to isolate the quintessence field $\phi$ from matter fields i.e. supersymmetric standard model (MSSM) fields and all other matter fields (BSM, moduli, dark matter) $\Phi$ is to write:
\begin{eqnarray}
K & = & K_{q}(\chi)+K_{m}(\Phi),\label{eq:K}\\
W & = & W_{q}(\chi)+W_{m}(\Phi).\label{eq:W}
\end{eqnarray}
Here $W_q(\chi),W_m(\Phi)$ are the quintessence and MSSM superpotentials. Since the K\"ahler metric is block diagonal we can take $\chi$ to be canonically normalised i.e. $K_q=\chi\bar{\chi}$ without affecting the other fields. To ensure that a quintessence potential of the form $V\sim\Lambda e^{-\beta\chi_{R}}$ (where $\Lambda$ is the cosmological constant today) is obtained we need $W_q=\frac{2}{\beta}\sqrt{\Lambda}e^{-\frac12 \beta\chi}$. 

The supergravity expression for the potential then gives a coupling between the MSSM potential and the quintessence field of the form (restoring $M_{\rm p}$):
\be
\delta V\sim\delta e^{K_q}|D_\Phi W|^{2}\sim\frac{\bar{\chi}_{0}}{M_{\rm p}^2}\delta\chi|D_\Phi W|^2.
\label{eq:qhiggs}
\ee
There is a similar coupling to fermions of the form: 
\be
\delta{\cal L}\sim\delta e^{K_{q}/2}{\cal L}_{{\rm fermion}}\sim\frac{\chi_0}{M_{\rm p}^2}\delta\chi \, {\cal L}_{{\rm fermion}} \,.
\label{eq:qfermi}
\ee

Also due to the Weyl anomaly there is a K\"ahler potential dependent correction to the physical (Einstein frame) gauge coupling:
\be
\frac{1}{g_{\rm phys}^2}=\Re f-\frac{3T(G)}{16\pi^2} \, K|_0,
\ee
(here $T(G)$ is a group theory number) giving similarly a quintessence dependence to the gauge field kinetic terms of the form:
\be
\delta{\cal L}\sim-\frac{3T(G)}{16\pi^{2}}\frac{\chi_{0}}{M_{\rm p}^2}\delta\chi \, {\cal {\cal L}}_{{\rm gauge}}.
\label{eq:qgauge}
\ee
By choosing the value of the quintessence field $\chi$ today to be $\chi_0=0$ one can of course avoid fifth-force bounds today. However, for a quintessence potential of the form $\hat{V}(\chi) = \Lambda^4 e^{-k \chi}$ with $k \simeq \mc{O}(1)$, the quintessence field is expected to move a distance $\Delta\phi\sim \mc{O}(1) M_{\rm p}$ and so at some point before the transition in the past to a matter/radiation dominated phase one would have had $\chi_{0}\sim \mc{O}(1) M_{\rm p}$. In this case we would have Planck suppressed linear couplings of the quintessence field for which there are strong bounds. 

Thus we have to conclude that at scales where the type of model such as that given in (\ref{Vin2}) is valid (and gives a quintessence model with no fifth-force issues) supersymmetry is badly broken. In other words the field content at these scales are not those of a supersymmetric theory\footnote{While this paper was being prepared for publication the paper \cite{Chiang:2018jdg} appeared which constructed two supergravity quintessence models. The first is an axionic model which we have argued would not be compatible with string theory expectations. The second model and its analysis appears to be consistent with our arguments above. We would also like to point out that the fact that a generic supergravity leads to a quintessence field with a mass equal to that of the gravitino can be avoided by having a term $m_q M_{\rm p} \chi$ in $W_q(\chi)$ and fine-tuning $m_q$ to cancel the term proportional to $m_{3/2}$.}.

\subsection{String quintessence models}
\label{sec:StringQuintessenceModels}

Candidates for quintessence fields can be naturally searched in the vast moduli sector of string compactifications. The first natural candidates that can be thought of are the overall volume and the dilaton since they are model independent. They may be rolling towards the infinite volume or zero coupling limit or towards a local minimum corresponding to dS or AdS. However they both couple to all matter fields and then the strong bounds on fifth-forces and varying constants would make this option untenable.

More promising models involve instead moduli which control the sizes of cycles hosting hidden sector branes and can in principle be a quintessence candidate. This is the case for the model considered in \cite{Cicoli:2012tz} which is based on an LVS string embedding with stabilised moduli \cite{Cicoli:2011yy} of the 6D Supersymmetric Large Extra Dimension (SLED) proposal \cite{Aghababaie:2003wz}. In this case low-scale gravity is used to address the hierarchy problem whereas supersymmetry is exploited to protect the quintessence potential from receiving dangerously large corrections. In fact, TeV-scale strings correlates with a gravitino mass of order the meV-scale. Notice that in these constructions supersymmetry is only non-linearly realised on the SM brane, and so $m_{3/2}$ is decoupled from the mass of the supersymmetric particles which is instead around the string scale. Moreover, in the presence of just two large extra dimensions, the Kaluza-Klein scale also reduces to the meV-scale. Hence corrections to the quintessence potential from loops of bulk fields of order $m_{3/2}^2 \Lambda_\cutoff^2$ scale as $({\rm meV})^4$, showing that supersymmetry is the symmetry which makes this quintessence model natural, similarly to the model of \cite{Albrecht:2001xt}. 

The difference between the two models is that in \cite{Albrecht:2001xt} the quintessence field is the overall volume mode which suffers from problems associated with the mediation of unobserved fifth-forces. On the other hand, in the string model of \cite{Cicoli:2012tz}, the quintessence field is a fibre modulus with a weaker-than-Planckian coupling to SM fields which does not give rise to any problem with fifth-forces. The reason for this small coupling is two-fold: ($i$) the SM is localised on a blow-up cycle which has no intersection with the fibre divisor and ($ii$) since the fibre divisor is a leading order flat direction, there is no leading order mixing between the quintessence field and the volume mode which can induce a Planck-strength coupling with SM fields. The main challenges of this model are the need to develop a proper 6D understanding since the gravitino mass is of order the Kaluza-Klein scale, together with the need to perform a detailed analysis of any correction which can induce a direct $M_{\rm p}$-suppressed coupling between the fibre divisor and SM fields. One should also carefully check that, according to the SLED proposal \cite{Aghababaie:2003wz}, loops of brane states are indeed cancelled by backreaction effects. 

Other quintessence candidates are the axionic partners of the moduli which have an approximate shift that protects their couplings \cite{Kaloper:2008qs,Panda:2010uq, Choi:1999xn}. Having compact support, the potential for these fields has to have a minimum. In order to perform a proper study of the implications for quintessence, they have to be considered within a full moduli stabilisation mechanism\footnote{See for instance the discussion in \cite{Panda:2010uq}.}. However, as mentioned above, one generically needs trans-Planckian axionic decay constants to drive a period of accelerated expansion \cite{Freese:1990rb} while recent studies of axion field ranges in string theory showed that this is very hard to achieve \cite{Klaewer:2016kiy, Blumenhagen:2017cxt, Palti:2017elp, Cicoli:2018tcq, Grimm:2018ohb, Heidenreich:2018kpg}.
 
A typical criticism of having a rolling scalar field in string compactifications with moduli stabilisation is that there is need for a double tuning, first the overall value of the cosmological constant and second the slope of the rolling field, both independently small numbers. However in LVS the axion partner of the volume modulus is a natural candidate for quintessence since its mass is of order $m\sim \mc{O}(e^{-\vo^{2/3}})$ and the volume exponentially large. For volumes of order $\vo\sim 10^3-10^5$ for which the string scale is of order GUT scale, the axion masses can be as small as $m\sim 10^{-32}$ eV  \cite{us}. In order to build a concrete model, one should check however that the axionic potential is flat enough to drive the present acceleration of our universe for a sub-Planckian decay constant. Furthermore any modulus that is stabilised by perturbative effects (by a combination of $\alpha'$ and loop effects) has a corresponding axionic partner with a mass which is doubly exponentially suppressed. It is then natural to have several candidates for dark energy but also for ultra-light axion dark matter. Notice that these fields can also give rise to condensates (axion stars) with masses of order $M_{\rm star}\sim M_{\rm p}^2/m$ which can be as heavy as $10^{20}$ solar masses \cite{Hui:2016ltb, Marsh:2015xka, Krippendorf:2018tei}. Moreover, as we shall show in Sec. \ref{sec:LowRedshift}, notice that ultra-light axions oscillating around a minimum with positive vacuum energy can induce an oscillating equation of state parameter which yields a small modification of the standard $\Lambda$CDM model.

\subsection{String axion quintessence}
\label{sec:LowRedshift}

We will consider now in more detail what in our opinion are the most promising candidates for a dynamical field for dark energy: string axions.

One of the most generic predictions of string theory is the existence of a \textit{string axiverse}~\cite{Svrcek:2006yi, Arvanitaki:2009fg, Cicoli:2012sz}, i.e. a large number of axions arising upon Kaluza-Klein reduction of the antisymmetric form fields on the internal cycles of the compactification space. The number of axions is related to the number of cycles in the compactification space and can easily be of $\mc{O}\left(100\right)$ or larger. Various non-perturbative corrections can give a mass to such fields, roughly of order $m_a \sim e^{-\tau}\,M_{\rm p}$, where $\tau$ is the corresponding saxion whose value parametrises the size of the cycle supporting non-perturbative effects. As $\tau$ can be rather large, i.e. much larger than the values needed to trust the effective field theory, the axion mass can easily be very small, as required for quintessence. Moreover, if the saxions receive a mass from perturbative effects, the low-energy EFT includes only the ultra-light axions. Concrete examples that feature all these properties are LVS models which admit at least one ultra-light axion corresponding to the axionic partner of the overall volume mode (which is stabilised perturbatively) with mass \cite{Conlon:2005ki, Cicoli:2017zbx}: 
\be 
m_a \simeq \sqrt{\frac{g_s}{8\pi}}\,\frac{M_{\rm p}}{\vo^{2/3}}\,e^{-\frac{\pi}{N} \vo^{2/3}} M_{\rm p} \,,
\ee
which can be in the right range for example for $g_s=0.1$, $N=3$ ($N$ is the rank of the condensing gauge group) and $\vo=1400$ as required to match the observed amplitude of the density perturbation in fibre inflation models \cite{Cicoli:2008gp}. LVS models with more than one large cycle would feature more ultra-light axionic candidates for explaining dark energy (as in the case of fibred CY threefolds where the fibre moduli are stabilised perturbatively and the corresponding axions remain light \cite{Cicoli:2008gp}). Another positive property of axions is that they feature a shift-symmetry at the perturbative level that naturally prevents their potential to acquire large quantum corrections. Finally ultra-light axions, being pseudo-scalars, can easily evade existing constraints from fifth-forces. For these reasons, axions are arguably one of the best candidate fields for quintessence in string theory. In this section we briefly review how axions can give rise to an accelerated late-time expansion of the universe.

In a moduli stabilisation scenario such as LVS we can separate the moduli between those that are stabilised by non-perturbative effects (such as blow-up modes) and those that are stabilised by perturbative effects (such as the overall volume and many fibre moduli). For the first group both the modulus and its corresponding axion get mass of the same order $m_a\sim m_{3/2}$. For the second group, the axions are much lighter than the moduli and we can study the EFT only for these ultra-light axions after integrating out all other massive fields. Since most known CY manifolds have a fibration structure, the number $N_\ULA$ of ultra-light axions can be very large ($N_\ULA\sim\mc{O}(100)$). To leading order in the non-perturbative expansion this axion potential takes the form\footnote{For potential generalisations of this scalar potential see for instance \cite{Bachlechner:2017zpb}.}:
\be
V=\Lambda^4-\sum_{i=1}^{N_\ULA} \Lambda_i^4 \cos\left(\frac{a_i}{f_i}\right) + \cdots \,,
\label{eq:AxionPotential}
\ee
where $f_i$ is the axion decay constant of the $i$-th canonically normalised axion field $a_i$, $\Lambda$ is the cosmological constant scale that can be tuned by fluxes and $\Lambda_i$ is the scale of the non-perturbative effect that gives mass to the $i$-th axion. In string compactifications the axion decay constant is roughly given by $f_i \simeq M_{\rm p}/ \tau_i < M_{\rm p}$ for $\tau_i >1$ \cite{Svrcek:2006yi, Arvanitaki:2009fg, Cicoli:2012sz}. For a quintessence candidate we need the slow-roll condition $\epsilon=\frac{M_{\rm p}^2}{2} \left(\frac{V'}{V}\right )^2 < 1$ to be satisfied. The scalar potential in \eqref{eq:AxionPotential} has a minimum at $\langle V\rangle = \Lambda^4-\sum_i\Lambda_i^4$ and a maximum at $V_{\rm max}=\Lambda^4+\sum_i\Lambda_i^4$ with  inflection points at $V_{\rm infl}=\Lambda^4$ as well as many ($2^{N_\ULA-1}$) saddle points. In phenomenological and cosmological discussions it is usually assumed that the minimum is tuned to zero but this is not natural in the landscape since the tuning for the overall minimum is not necessarily related with the scales of each the $\Lambda_i$'s. Therefore we may study different possibilities in particular for $\Lambda$ greater, smaller or of the same order as the smallest $\Lambda_i$.

Depending on the values of $\tau_i$ and the coefficients of the non-perturbative effects, the corresponding axions can also be integrated out until we reach the lightest one, that we denote with $a_\ell$. Focusing for simplicity on $a_\ell$, the corresponding slow-roll condition is:
\be
\epsilon = \frac12 \left[\left(\frac{\Lambda_\ell}{\Lambda}\right)^4\frac{M_{\rm p}}{f_{\ell}}\right]^2 \frac{\sin^2\left(a_\ell/f_{\ell}\right)}{\left(1 - \left(\Lambda_{\ell}/\Lambda\right)^4 \cos\left(a_\ell/f_\ell\right)\right)^2} < 1 \,.
\label{eq:epsilon}
\ee
However, before integrating out the heavier axions, the original potential can give rise to interesting early universe cosmology. In particular, as the universe evolves and the Hubble parameter decreases, each axion field is essentially frozen at its value after inflation due to the large Hubble friction. Once the Hubble scale hits the mass threshold of a given axion, the axion  starts to roll and oscillates around its minimum. Depending on the relative values of $\Lambda$ and $\Lambda_i$ as well as the initial value of the field, the slow roll condition may or may not be satisfied.

Depending on the values of different constants we will have distinctive scenarios which we now state:
\ben
\item \textbf{Alignment mechanism}: If the minimum of the potential is tuned to be at vanishing energy (i.e. if $\Lambda = \Lambda_\ell$) as is usually done in the literature, we can observe from eq. \eqref{eq:epsilon} that in order to get an accelerated expansion of the universe the axion decay constant has to be $f_\ell \gtrsim M_{\rm p}$. Getting a (super-)Planckian axion decay constant is a well-known issue in string theory since it is in tension with the fact that the cycles volumes are expected to be larger than the string scale ($\tau_\ell\gtrsim 1$). However there might be possible way-outs that rely on alignment mechanisms involving two~\cite{Kim:2004rp, Shiu:2015xda} or many fields~\cite{Dimopoulos:2005ac, Cicoli:2014sva}.

\item \textbf{Hilltop quintessence}: As explained above, the generic situation is to have axions with sub-Planckian decay constants. In this case, even if $\Lambda = \Lambda_\ell$, the axion $a_\ell$ could still drive the present epoch of accelerated expansion without the need to rely on complicated misalignment-like mechanisms. In fact, if the maximum of the potential for $a_\ell$ is located at positive energy (i.e. $\Lambda^4 + \Lambda_i^4 > 0$), as in the two examples reported in Fig. \ref{fig:potentials}, and the field is initially displaced close to it, the universe undergoes accelerated expansion~\cite{Kamionkowski:2014zda}. Notice that in order for this mechanism to work, the minimum of the potential does not need to be tuned to $0$: the crucial point is just that a region of the potential around the maximum is at positive energy. Moreover, axion fields are very light, and so it is very easy to displace them from their minima, e.g. during inflation. Given the large number of ultra-light axions in generic string compactifications, we expect that the displacement of these fields is evenly distributed in the range $a_i/f_i \in [-\pi,\pi]$, and so it should not be difficult to find one of them around its maximum. We stress that this case would be the only way to get axion inflation when $f_\ell < M_{\rm p}$ even if the initial position of the axion has to be tuned extremely close to the maximum to obtain enough efoldings of inflation \cite{BlancoPillado:2004ns}.

\item \textbf{Quasi-natural quintessence}: Notice that in the landscape there is no reason to tune the minimum to vanishing vacuum energy. If the minimum of the potential for the lightest axion is tuned to be of the order of the current value of the cosmological constant $\Lambda$, the slow-roll condition just implies (the term which depends on $a_\ell/f_\ell$ in eq. \eqref{eq:epsilon} is always smaller than $1$):
\be
f_{\ell}\gtrsim \left(\frac{\Lambda_\ell}{\Lambda}\right)^4\,M_{\rm p}\,,
\label{eq:IPQ}
\ee
which allows for a sub-Planckian axion decay constant $f_\ell<M_{\rm p}$ as long as $\Lambda \gg \Lambda_\ell$. The slow-roll condition $\epsilon<1$ is naturally satisfied for a very large region of field space, not only close to the hilltop as can be seen in Fig. \ref{fig:IPQ}). The corresponding equation of state would give a small modification to the cosmological constant scenario:
\be
w=\frac{p}{\rho}=\frac{\frac{{\dot a}^2}{2}-V}{\frac{{\dot a}^2}{2}+V}\sim -\frac{1-\frac13 \epsilon}{1+\frac13 \epsilon}\sim -1 + \frac{2}{3}\epsilon \,.
\ee
It is worth mentioning that the case $\Lambda \gg \Lambda_\ell$ is never considered for inflation since the energy scale of the potential would be of order the cosmological constant scale, and so would be way too low to match the observed amplitude of the density perturbations. Moreover, for $\Lambda \gg \Lambda_\ell$, if $f_\ell$ is not too low, $\epsilon$ is below unity everywhere in the axion field space, and so there would be no way to end inflation.

\item \textbf{Oscillating scalar}: Another possible modification of the constant dark energy scenario could be given by an oscillating axion. Assuming that $\Lambda$ is tuned at the current value of the cosmological constant as in the left panel of Fig. \ref{fig:Oscillating-wt} and that $a_\ell$ is initially displaced from its minimum, the field starts oscillating around its minimum when $H$ is of order of its mass. This will then produce an interesting oscillating equation of state, as shown in the right panel of Fig. \ref{fig:Oscillating-wt} for $f/M_{\rm p} = 1$ and $\Lambda_\ell/\Lambda =0.85$. This expected behaviour could be used to study the existence of axions with mass of the same order of $H_0$, comparing with low-redshift observations.
\een

\begin{figure}
\begin{center}
\includegraphics[width=\textwidth]{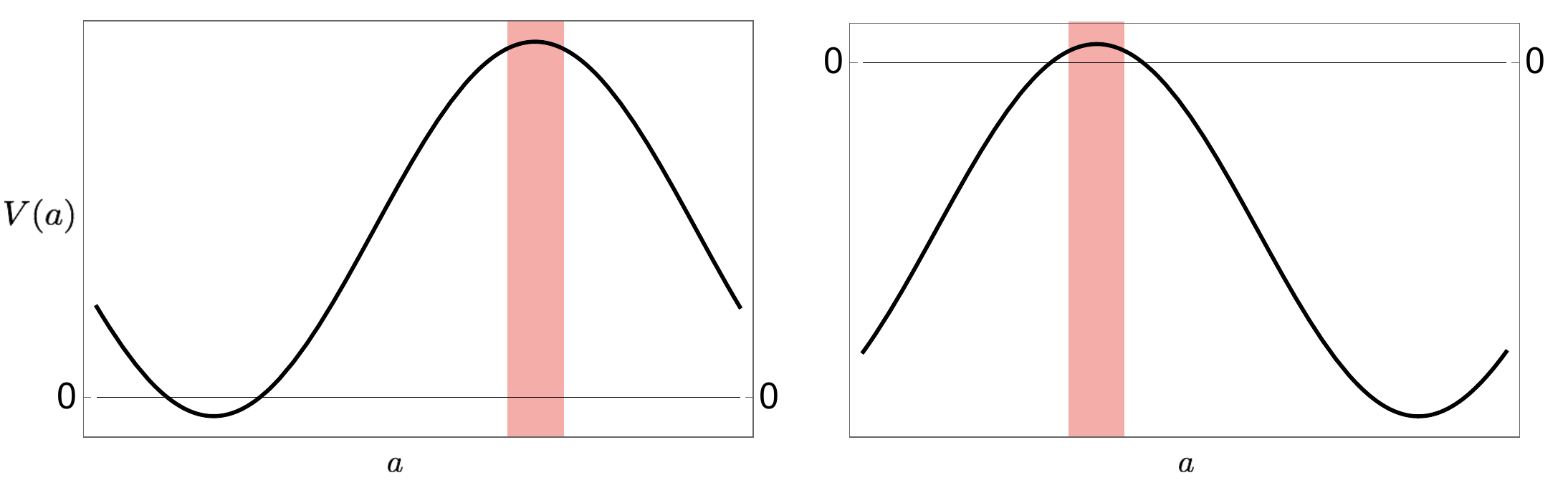}
\end{center}
\caption{Examples of potentials that allow for hilltop quintessence. The red domains schematically represent the regions of the potentials where slow-roll can take place.}
\label{fig:potentials}
\end{figure}

\begin{figure}
\begin{center}
\includegraphics[width=0.7\textwidth]{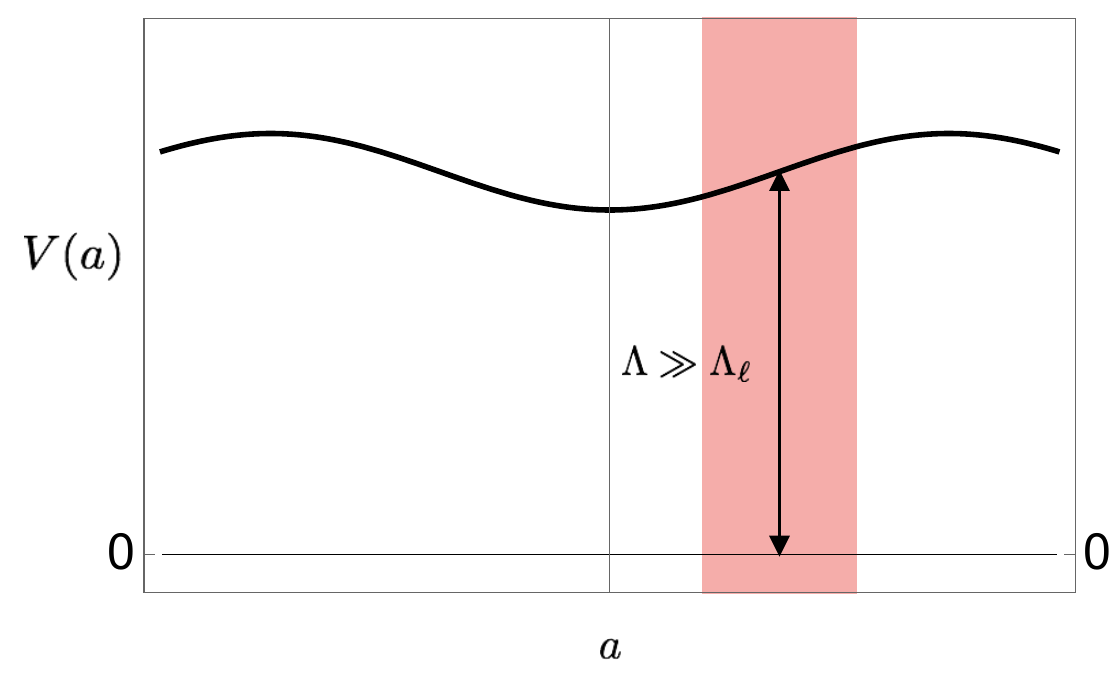}
\end{center}
\caption{In the case $\Lambda \gg \Lambda_\ell$ slow-roll can happen also in the region close to the inflection point of the potential, and given (\ref{eq:IPQ}) this does not require a super-Planckian axion decay constant.}
\label{fig:IPQ}
\end{figure}

\begin{figure}
\begin{center}
\includegraphics[width=\textwidth]{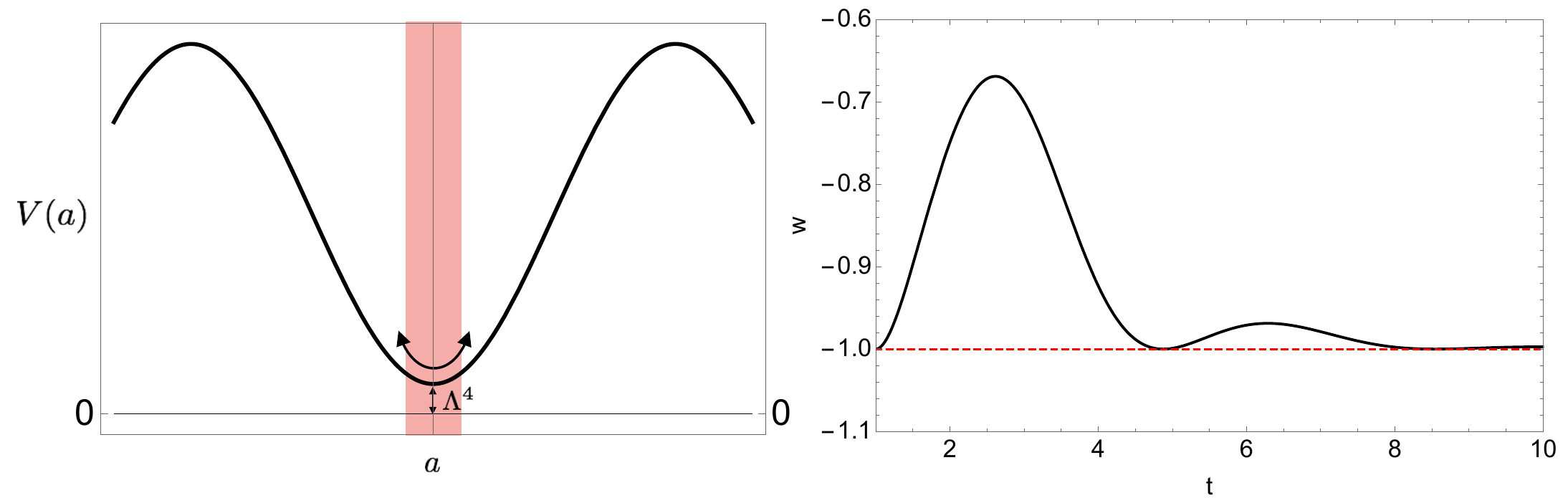}
\end{center}
\caption{We illustrate how the equation of state oscillates while the axion oscillates around its minimum (time is in units of the axion mass). Contrary to the dark matter case in which the average $w$ vanishes, here the presence of $\Lambda$ causes the average to be non-zero. This behaviour can be compared with data from low-redshift observations, in order to explore the existence of axions with mass around $H_0$.}
\label{fig:Oscillating-wt}
\end{figure}

Notice that cases ($3$) and ($4$) necessarily violate the swampland conjecture (\ref{eq:SwamplandConjecture}) since they require dS minima, while case ($1$) would violate the swampland conjecture on field distances \cite{Ooguri:2006in} since it requires trans-Planckian physics. On the other hand, as shown in Fig. \ref{fig:potentials}, case ($2$) just requires the presence of a maximum at positive energy but it would work also for sub-Planckian axion decay constants. Hence this case would violate the swampland conjecture (\ref{eq:SwamplandConjecture}) but it would still be allowed by a refined conjecture which does not exclude dS maxima \cite{Andriot:2018wzk}.

Considerations of ultra-light axions corresponding to a quintessence field have been made in several recent studies \cite{Kamionkowski:2014zda, Karwal:2016vyq,Poulin:2018dzj}. The fact that there may be many axions dominating the energy density at different stages of the evolution of the universe may be a way to address the apparent discrepancy among the different measurements of $H$ at high and low redshift. Their considerations can be adapted to the present discussion but with the difference that we do not assume the minimum of the potential to vanish. 

In summary string theory axions provide  interesting candidates to be quintessence for several reasons:
\bi
\item Ultra-light axions are a natural outcome of moduli stabilisation scenarios with exponentially suppressed masses. 

\item Depending on the value of these masses, the axions can be ultra light dark matter or dark energy.

\item These ultra-light axions are also natural candidates for dark radiation produced after the decay of the corresponding modulus field \cite{Cicoli:2012aq, Higaki:2012ar, Angus:2014bia, Hebecker:2014gka, Cicoli:2015bpq} which can put constraints on string scenarios but also can partially address cosmological issues such as the tension between high and low redshift measurements of the Hubble parameter by increasing the value of $N_{\rm eff}$ \cite{Riess:2016jrr}\footnote{Notice however that larger values of $N_{\rm eff}$, even if they decrease the tension between different determinations of $H_0$, increase the existing tension between different measurements of the $\sigma_8$ parameter.}.

\item The fact that there may be hundreds or thousands of ultra-light axions can give rise to interesting cosmological periods in early universe cosmology with also potential implications for different measurements of $H$.

\item If the overall minimum of the potential is not tuned at zero several scenarios emerge with accelerating universes. A negative vacuum energy is allowed if slow-roll starts close to a maximum or a saddle point at positive $V$ and the slow-roll condition can be easily satisfied with no trans-Planckian decay constant as long as $\Lambda \gg \Lambda_\ell$. The different axions oscillating around their minima do not risk overclosing the universe since the minimum is not at zero. An oscillating scalar around a minimum with positive vacuum energy can give rise to a varying equation of state. The time in which the field climbs the potential may mimic $w<-1$ as suggested in \cite{Csaki:2005vq}. However, reproducing the recent analysis, which suggests a turning point for the Hubble parameter \cite{Wang:2018fng, Capozziello:2018jya, Dutta:2018vmq}, remains a theoretical challenge if these results were confirmed.
\ei

\section{Conclusions}

In this paper we have analysed general aspects regarding dS and quintessence scenarios to have a concrete realisation in effective field theories derived from string compactifications. 

We have seen that even though in order to have full control of dS moduli stabilisation a non-perturbative formulation of string theory is needed, there has been substantial progress in the past decades to be confident that these solutions do exist and that the string theory landscape is a generic outcome of string theory. It is actually remarkable that, without having a full non-perturbative formulation of the theory and not knowing even the metric of the extra dimensional manifolds, there is a coherent picture in which all moduli are stabilised and dS space in 4D can appear as a solution. 

It is worth emphasising that this procedure uses explicit string theory features with solid mathematical structures such as the topological properties of the compact space, warping induced by fluxes, tadpole cancellation conditions, brane and anti-brane dynamics, explicit computations of leading order perturbative and non-perturbative corrections to the effective field theory, etc. It is fair to say that a full control is difficult to achieve with our current understanding of string compactifications which are not  maximally supersymmetric but not having full control on the calculations should not be confused with  having no control at all. The results are based on well defined approximations which are justified as long as the couplings are weak and the volumes are large enough. Luckily this is the regime that is also interesting for phenomenological applications\footnote{Notice that the challenge to obtain proper inflationary models from string theory with large tensor modes is mostly due to the fact that, if these modes were observable, the corresponding EFT would be at the edge of its validity.}.

We have also seen that the natural alternative to dS space, quintessence, can also be accommodated in string compactifications albeit in a more complicated way. Having a rolling direction which is flat enough to give rise to the observed dark energy requires all other moduli to be stabilised in a similar way as in dS compactifications or rolling even more slowly, something which is more challenging than getting dS. Typical candidates for the quintessence field such as the overall volume and the dilaton are not appropriate to be the quintessence field since they couple to all matter in hidden and observable sectors, and so would be subject to stringent fifth-force constraints \cite{Adelberger:2003zx}. On the other hand, moduli associated with cycles hosting only hidden sector fields \cite{Cicoli:2012tz} may still be allowed by observational constraints although that may require a very small string scale. A low string scale has also appeared in efforts to construct quintessence models in warped throats \cite{Panda:2010uq}.

We also studied the nature of the Higgs couplings to various fields in light of the swampland conjectures. We have found that a direct coupling between the Higgs and the quintessence field can be avoided if there are other fields which give non-trivial contributions to $\nabla V$ at the symmetric point of the Higgs potential. However such realisations seem difficult to realise from the point of view of string theory. 

We have also analysed the possibility of quintessence in the context of supergravity and illustrated the presence of generic couplings (including one loop effects) between all fermions and the quintessence field, which are in tension with the observational bounds. Moreover, analysis of renormalisation group effects showed the requirement of functional fine-tuning of the tree-level potential of the quintessence field or at least additional fine-tuning compared to dS models.

The best candidates of quintessence fields are the multiple axions that abound in string compactifications. Considering them just as rolling quintessence fields is a very limited option. However since they correspond to periodic fields with a compact support, their scalar potential has to have a minimum. A natural possibility is that these fields may be oscillating around a dS minimum giving rise to a small modification of the standard $\Lambda$CDM scenario.  

It is important to notice that in general some of the K\"ahler moduli obtain mass via perturbative effects. This implies that the corresponding axions, which get lifted only by non-perturbative effects, are much lighter yielding a large mass hierarchy among the two components of the same complex scalar field. This is precisely the case for the overall volume modulus and  fibration moduli in LVS models. It is worth emphasising that having an extremely light axion is the most model-independent prediction of LVS constructions. Having fibre moduli is also very generic. It is then possible to have one of the axions to correspond to ultra light dark matter with a mass of order $10^{-22}$ eV and another to provide dark energy with a mass of order $10^{-32}$ eV. Furthermore both can be candidates to be part of dark radiation for which there are strong constraints. This justifies a more detailed study of the cosmological implications of these light axions.

It would be highly desirable to count on a non-perturbative formulation of string theory that could hopefully determine once and for all that there are or not dS or quintessence solutions of string theory (as it would also be good to have a full proof of the AdS/CFT correspondence or the finiteness of string theory or any potential alternative). As usual in science we have to content ourselves to extract information based on limited experimental input  and theoretical control. In the Standard Model we have experimental data which we can confront whereas string constructions cannot at the moment be discriminated on the basis of observations. In the case of dS vs quintessence we hope we have argued that the theoretical progress made over the years, although not $100\%$ satisfactory, is encouraging and present a coherent picture. Furthermore, experimentally, the fact that the equation of state $w$ has been converging over the years towards $w=-1$ is tantalising to bend the preference in favour of dS, following standard Bayesian criteria. However, the recent tension among values of the Hubble parameter determined from high and low redshift may hint at a variable equation of state that could be at odds with both dS and quintessence (see for instance \cite{Wang:2018fng, Capozziello:2018jya, Dutta:2018vmq}). Even though it is too early to judge the robustness of this analysis we have to keep an open mind. Low redshift measurements have surprised us already once, against our theoretical prejudices, and may do it again. 
 
Finally we would like to remark that it is healthy to challenge the different approaches to obtain dS space in a fundamental theory. Having criticism and skepticism to a concrete scientific development helps to sharpen the arguments and clarify the achievements and open questions. In view of the lack of further experimental input, this is the best avenue to address theoretical questions and converge towards the best possible explanations. In the case of dS vacua, the question is of utmost importance and  having an open debate helps to streamline all the arguments and eventually improve the existing constructions to make them more explicit and coherent or even rule them out. Given the importance of the question being addressed, a high level of scrutiny of the solutions is important -- in fact no bar is too high a bar. We hope to come back and address some of the open questions highlighted in this article.

\section*{Acknowledgements}

The authors would like to thank B. Acharya, D. Andriot, C.P. Burgess, J.P. Conlon, O. DeWolfe, D. Ghosh, A. Hebecker, S. Krippendorf, L. Pando-Zayas, V. Poulin, A. A. Sen, A. Sen, R. Sheth, C. Vafa, R. Valandro and G. Villadoro for discussions. We also want to acknowledge discussions at different stages with  J. Moritz,  J. Polchinski, A. Retolaza,  R. Savelli, S. Sethi, G. Shiu and A. Westphal. MC, SdA and AM would like to thank ICTP for hospitality. AM is supported in part by a Ramanujan Fellowship, DST, Government of India.

\appendix

\appendix

\section{Checking the swampland conjecture for model 2}

Now, we turn to explicitly verifying that the swampland criterion is not violated by the potential for model 2. For concreteness we will take the potential of the field $\phi$ to be $g(\phi) = \mu^2 \tilde{\Lambda}^2 \ln \cosh ( {\phi  \over \tilde{\Lambda}} )$ and $\Lambda \sim \mathcal{O}(\mu)$. The quintessence contribution becomes important only around the critical point with $h = v$ and $\chi = 0$ where the energy density stored in the two fields $h$ and $\chi$ vanishes. There, the conjecture gives $\Psi \equiv \left|\nabla V\right| \big{/} V  \simeq |\beta|$ and $\beta$ can be of order $\mc{O}(1)$ for quintessence.

It is easy to realise (and check numerically) that the conjecture is satisfied for all the regions in field space away from the critical points of the potential. Since we are interested in the field range $|\phi| \lesssim 1$, the exponentials can be estimated with numbers of $\mathcal{O}(1)$: $e^{{\phi}} \sim 1$. Recall  that we are interested in the region $\mu \ll v$. For instance, consider the region ${v} \ll {h} \ll 1$. In this region, the potential and its derivatives behave parametrically as:
\begin{align}
|\nabla V| &\sim \mc{O}(h^3) + \left(\mc{O}(h^4) + \mc{O}(\mu^3)\right) \simeq \mc{O}(h^3) \,, \\
V &\sim \mc{O}(h^4) + \mc{O}(\mu^4) \sim \mc{O}(h^4) \,,
\end{align}
where we have used the fact that $\mu \ll v \ll h$. The first term in $|\nabla {V}|$ comes from the Higgs potential, while the terms in bracket come from ${V}_{{\phi}}$, analogously considerations hold for the potential V. This gives ${\Psi} \sim \mc{O}(1/{h}) \gtrsim 1$ for $h \lesssim 1$. Similar arguments hold in the region $0 \ll {h} \ll {v}$. For example, restricting to the region $\mu \ll h \ll v$. Then the potential and its derivatives behave parametrically as\footnote{The term $\mc{O}(\mu^3)$ is subleading with respect to $\mc{O}({h} {v}^2)$. The result depends on whether $h \ll v^2$ or the opposite. In the former case $|\nabla {V}| \simeq \mc{O}({v}^4)$, otherwise $|\nabla {V}| \simeq \mathcal{O}({h} {v}^2)$. In both cases the function $\Psi$ is bounded from below.}
\begin{align}
|\nabla {V}| &\sim \mc{O}({h} {v}^2) + \left(\mc{O}({v}^4) + \mc{O}({\mu}^3)\right) \,, \\
{V} &\sim \mc{O}({v}^4) + \mc{O}({\mu}^4) \sim \mc{O}({v}^4) \,,
\end{align}
then the function ${\Psi}$ is bounded from below and does not fall below unity. The region ${h} \ll {\mu} \ll {v}$ can be examined similarly.

\begin{figure}
\begin{center}
\includegraphics[width=0.6\textwidth]{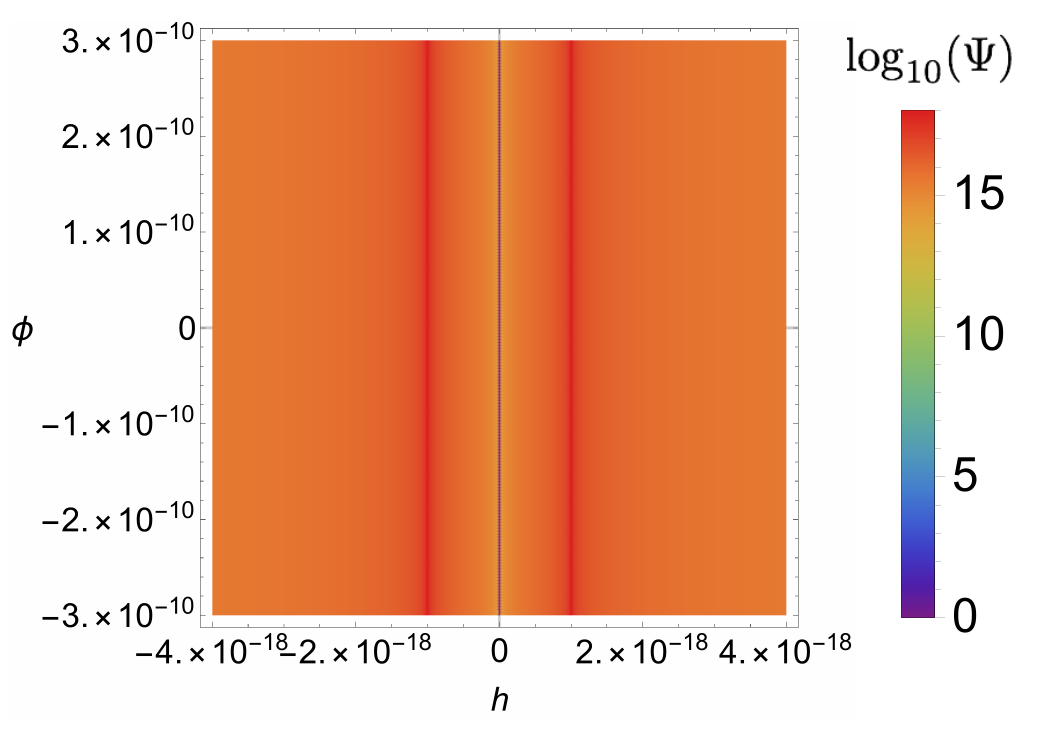}
\end{center}
\caption{We plot the function $\Psi$ in the $\chi$ direction at $h = v$.}
\label{fig:DensityPlot}
\end{figure}

Concerning the critical points, in the broken Higgs vacuum ${h} = {v}$ and for ${\chi} = 0$ the conjecture is trivially satisfied due to the contributions from the quintessence field. The behaviour of the function in a neighborhood of this critical point is shown in Fig. \ref{fig:DensityPlot}. The symmetric point of the Higgs potential ($h = 0$) does not violate the conjecture, contrary to what happens in the two-field case of~\cite{Denef:2018etk}. As shown in Fig. \ref{fig:DensityPlot}, the direction $h = 0$ is potentially dangerous. However, it is immediate to see that the function $\Psi$ is bounded from below in that locus: $\Psi \gtrsim 1$. Another interesting locus to examine is that of $V_{\phi} = 0$ (for values the of $h$ for which solutions to this exists). For values of $h$ such that $V(h) \ll  \mu^3$, we have $\phi \sim \epsilon \mu$ where $\epsilon \equiv V(h)/  \mu^3 $. With this, the contributions to the potential and the its gradient from the field $\phi$ are subdominant  and $\Psi \sim {1 \over \sqrt{\epsilon}}$. On the other hand, for $V(h) \sim 
\mu^3$, $\phi \gtrsim \mu$. In this regime, $|\nabla V| \sim  v \mu^{3/2}$ and there is no violation of the conjecture. Similar arguments hold for $\phi \sim \mu$.

\end{document}